\documentclass[11pt,a4paper]{article}

\usepackage{amssymb}
\usepackage[dvips]{graphicx}
\usepackage{bm}

\begin{document}

\title{\bf Three-loop NSVZ relation for terms quartic in the Yukawa couplings with the higher covariant derivative regularization}

\author{
V.Yu.Shakhmanov, K.V.Stepanyantz\\
{\small{\em Moscow State University}}, {\small{\em Faculty of Physics, Department  of Theoretical Physics}}\\
{\small{\em 119991, Moscow, Russia}}}

\maketitle

\begin{abstract}
We demonstrate that in non-Abelian ${\cal N}=1$ supersymmetric gauge theories the NSVZ relation is valid for terms quartic in the Yukawa couplings independently of the subtraction scheme if the renormalization group functions are defined in terms of the bare couplings and the theory is regularized by higher covariant derivatives. The terms quartic in the Yukawa couplings appear in the three-loop $\beta$-function and in the two-loop anomalous dimension of the matter superfields. We have obtained that the three-loop contribution to the $\beta$-function quartic in the Yukawa couplings is given by an integral of double total derivatives. Consequently, one of the loop integrals can be taken and the three-loop contribution to the $\beta$-function is reduced to the two-loop contribution to the anomalous dimension. The remaining loop integrals have been calculated for the simplest form of the higher derivative regularizing term. Then we construct the renormalization group functions defined in terms of the renormalized couplings. In the considered approximation they do not satisfy the NSVZ relation for a general renormalization prescription. However, we verify that the recently proposed boundary conditions defining the NSVZ scheme in the non-Abelian case really lead to the NSVZ relation between the terms of the considered structure.
\end{abstract}

\unitlength=1cm

\section{Introduction}
\hspace*{\parindent}

The exact NSVZ $\beta$-function \cite{Novikov:1983uc,Jones:1983ip,Novikov:1985rd,Shifman:1986zi} is the equation which relates the $\beta$-function of ${\cal N}=1$ supersymmetric gauge theories to the anomalous dimensions of the matter superfields $(\gamma_\phi)_j{}^i$ and gives the exact $\beta$-function for the pure ${\cal N}=1$ supersymmetric Yang--Mills (SYM) theory,\footnote{Note that so far we do not specify the definitions of the renormalization group functions. They will be discussed in details later.}

\begin{equation}\label{NSVZ}
\beta = - \frac{\alpha^2\Big(3 C_2 - T(R) + C(R)_i{}^j (\gamma_\phi)_j{}^i/r\Big)}{2\pi(1-C_2\alpha/2\pi)}.
\end{equation}

\noindent
Here $r=\delta_{AA}$ is the dimension of the gauge group; $C(R)_i{}^j\equiv (T^A T^A)_i{}^j$ with $T^A$ being the generators of the gauge group in the representation to which the chiral matter superfields belong. $T(R)$ is defined by the equation $\mbox{tr}(T^A T^B)\equiv T(R)\delta^{AB}$, and $C_2 = T(Adj)$.

The NSVZ relation is closely connected with the ${\cal N}=2$ non-renormalization theorem \cite{Shifman:1999mv,Buchbinder:2014wra,Buchbinder:2015eva} (which states that the divergences in ${\cal N}=2$ SYM theories exist only in the one-loop approximation \cite{Grisaru:1982zh,Howe:1983sr,Buchbinder:1997ib}). There are also NSVZ-like equations in the softly broken ${\cal N}=1$ supersymmetric theories \cite{Hisano:1997ua,Jack:1997pa,Avdeev:1997vx}.

Originally the NSVZ relation has been obtained from various general arguments based, e.g., on the structure of instanton contributions \cite{Novikov:1983uc,Novikov:1985rd,Shifman:1999mv}, anomalies \cite{Jones:1983ip,Shifman:1986zi,ArkaniHamed:1997mj}, non-renormalization of the topological term \cite{Kraus:2002nu}. However, straightforward perturbative calculations indicate that the NSVZ relation is not valid in the $\overline{\mbox{DR}}$ subtraction scheme \cite{Jack:1996vg,Jack:1996cn,Jack:1998uj} and in the MOM subtraction scheme \cite{Kataev:2013csa}. This is caused by the scheme dependence of the NSVZ relation \cite{Kutasov:2004xu,Kataev:2014gxa}. The NSVZ scheme can be related to the above mentioned schemes by finite renormalizations \cite{Jack:1996vg,Jack:1996cn,Jack:1998uj,Aleshin:2016rrr}. Note that the possibility of making these finite renormalizations is highly non-trivial, because the NSVZ relation leads to some scheme independent consequences \cite{Kataev:2013csa,Kataev:2014gxa}. Nevertheless, in the case of using the dimensional reduction the NSVZ scheme should be tuned in each order of the perturbation theory, and there is no simple prescription giving it in all orders (see, e.g., \cite{Aleshin:2016rrr}). Such a prescription \cite{Kataev:2013eta} can be given in the case of using the Slavnov higher derivative regularization \cite{Slavnov:1971aw,Slavnov:1972sq,Slavnov:1977zf} in the supersymmetric version \cite{Krivoshchekov:1978xg,West:1985jx}. Presumably, with the higher derivative regularization the renormalization group (RG) functions defined in terms of the bare coupling constant satisfy the NSVZ relation in all orders independently of the subtraction scheme. This occurs because the $\beta$-function seems to be determined by integrals of double total derivatives.\footnote{In the case of using the dimensional reduction \cite{Siegel:1979wq} such a factorization does not take place \cite{Aleshin:2015qqc}.} The factorization into integrals of total derivatives and double total derivatives has first been noted in \cite{Soloshenko:2003nc} and \cite{Smilga:2004zr}, respectively. Subsequently, for various supersymmetric theories it has been verified by numerous calculations in the lowest orders of the perturbation theory \cite{Shevtsova2009,Pimenov:2009hv,Stepanyantz:2011bz,Stepanyantz:2012zz,Stepanyantz:2012us,Kazantsev:2014yna,Buchbinder:2014wra,Buchbinder:2015eva} and even proved in all orders in the Abelian case \cite{Stepanyantz:2011jy,Stepanyantz:2014ima}. Similar factorizations into integrals of double total derivative have been proved in orders for the Adler $D$-function \cite{Adler:1974gd} in ${\cal N}=1$ SQCD \cite{Shifman:2014cya,Shifman:2015doa} and for the anomalous dimension of the photino mass in the softly broken ${\cal N}=1$ SQED \cite{Nartsev:2016nym}. In both cases they allow all-order proving of the NSVZ-like relations for the RG functions defined in terms of the bare coupling constant.

For the scheme-dependent RG functions (standardly, \cite{Bogolyubov:1980nc}) defined in terms of the renormalized coupling constant the NSVZ scheme can be obtained in all orders in the Abelian case by imposing simple boundary conditions to the renormalization constants \cite{Kataev:2013csa,Kataev:2014gxa,Kataev:2013eta}. The NSVZ scheme for the photino mass anomalous dimension has been constructed by this method in \cite{Nartsev:2016mvn}.

For non-Abelian gauge theories, regularized by higher derivatives, the NSVZ relation for the RG functions defined in terms of the bare couplings has not yet been derived by the tools of the perturbation theory. However, at the qualitative level, the appearance of the NSVZ $\beta$-function has been explained in \cite{Stepanyantz:2016gtk}, where the NSVZ equation was rewritten as a relation between the $\beta$-function and the anomalous dimensions of the quantum gauge superfield, of the Faddeev--Popov ghosts, and of the matter superfields. This allows to suggest that for the higher covariant derivative regularization in the non-Abelian case the NSVZ relation is also valid for the RG functions defined in terms of the bare couplings and has the form

\begin{equation}\label{New_NSVZ}
\frac{\beta(\alpha_0,\lambda_0)}{\alpha_0^2} = - \frac{1}{2\pi}\Big(3 C_2 - T(R) - 2C_2 \gamma_c(\alpha_0,\lambda_0) - 2C_2 \gamma_V(\alpha_0,\lambda_0) + C(R)_i{}^j \gamma_\phi(\alpha_0,\lambda_0)_j{}^i/r\Big).
\end{equation}

\noindent
Consequently, the prescription giving the NSVZ scheme for the RG functions defined in terms of the renormalized couplings in the non-Abelian case is

\begin{equation}\label{Scheme_Prescription}
Z_\alpha(\alpha,\lambda,x_0) = 1;\qquad Z_\phi(\alpha,\lambda,x_0)_i{}^j=\delta_i{}^j;\qquad Z_c(\alpha,\lambda,x_0)=1; \qquad Z_V = Z_\alpha^{1/2} Z_c^{-1},
\end{equation}

\noindent
where $x_0$ is a fixed value of $x=\ln\Lambda/\mu$ with $\Lambda$ and $\mu$ being a dimensionful parameter of the regularized theory and a normalization point, respectively.

Certainly, it is necessary to verify these statements by explicit perturbative calculations. Taking into account that the $\beta$-function is scheme-dependent starting from three loops, and the anomalous dimensions are scheme-dependent starting from two loops, for non-trivial checking of the above statements one has to compare the three-loop $\beta$-function with the two-loop anomalous dimension. The complete three-loop calculation is rather complicated, so that in this paper we consider only a part of it. Namely, we consider only the terms quartic in the Yukawa couplings. The purpose of this paper is to verify that the $\beta$-function is given by integrals of double total derivatives and check Eqs. (\ref{New_NSVZ}) and (\ref{Scheme_Prescription}) for the terms of this structure.

The paper is organized as follows. In Sect. \ref{Section_Theory} we consider the ${\cal N} =1$ SYM theory with matter superfields regularized by higher derivatives and introduce the notation. The supergraphs defining the terms quartic in the Yukawa couplings in the three-loop $\beta$-function and in the two-loop anomalous dimension are calculated in Sect. \ref{Section_Calculation}. In particular, in this section we demonstrate that the considered contribution to the $\beta$-function can be presented as an integral of a double total derivative in the momentum space. Moreover, we obtain that the considered parts of the RG functions defined in terms of the bare couplings satisfy the NSVZ relation independently of the subtraction scheme with the higher covariant derivative regularization. In Sect. \ref{Section_Explicit_Gamma} for the simplest form of the higher derivative regulator we calculate the integrals giving the part of the two-loop anomalous dimension quartic in the Yukawa couplings. The explicit expression for the anomalous dimension obtained in Sect. \ref{Section_Explicit_Gamma} is used in Sect. \ref{Section_NSVZ} for checking the prescription (\ref{Scheme_Prescription}) which gives the NSVZ scheme for the RG functions defined in terms of the renormalized couplings. In particular, we calculate the considered parts of the RG functions defined in terms of the renormalized couplings. One can see that the part of the anomalous dimension quartic in the Yukawa couplings is scheme independent and coincides with the result obtained earlier in the $\overline{\mbox{DR}}$ scheme (see \cite{Jack:1996vg} and references therein), while the part of the $\beta$-function quartic in the Yukawa couplings is scheme dependent. Then we demonstrate that under the prescription (\ref{Scheme_Prescription}) the NSVZ relation is really valid for the considered contributions to the RG functions (defined in terms of the renormalized couplings). In the Appendixes we present explicit expressions for individual superdiagrams and describe in details the calculation of the loop integrals.

\section{The ${\cal N}=1$ SYM theory regularized by higher derivatives}
\hspace*{\parindent}\label{Section_Theory}

In this paper we will consider the general ${\cal N}=1$ SYM theory with matter in the massless limit. In terms of superfields \cite{West:1990tg,Buchbinder:1998qv} it is described by the manifestly supesymmetric action

\begin{eqnarray}\label{SYM_Action_Original}
&& S = \frac{1}{2 e_0^2}\,\mbox{Re}\,\mbox{tr}\int d^4x\,
d^2\theta\,W^a W_a + \frac{1}{4} \int d^4x\, d^4\theta\,\phi^{*i}
(e^{2V})_i{}^j \phi_j\nonumber\\
&&\qquad\qquad\qquad\qquad\qquad\qquad\qquad\qquad\qquad  +
\Big(\frac{1}{6} \int d^4x\,d^2\theta\,\lambda_0^{ijk} \phi_i
\phi_j \phi_k + \mbox{c.c.}\Big),\qquad
\end{eqnarray}

\noindent
where $V$ is a real gauge superfield and $\phi_i$ are chiral matter superfields in a certain representation $R$ of the gauge group $G$. The supersymmetric gauge field strength $W_a = \bar D^2 (e^{-2V} D_a e^{2V})/8$ is also a chiral superfield; $e_0$ and $\lambda_0^{ijk}$ are the bare gauge and Yukawa couplings, respectively. We assume that the theory is gauge invariant, so that

\begin{equation}\label{Yukawa_Identity}
\lambda_0^{ijm} (T^A)_m{}^{k} + \lambda_0^{imk} (T^A)_m{}^{j} +
\lambda_0^{mjk} (T^A)_m{}^{i} = 0,
\end{equation}

\noindent
where $(T^A)_i{}^j$ are the generators of the representation $R$. The generators of the fundamental representation are denoted by $t^A$. By definition, they satisfy the normalization condition $\mbox{tr}(t^A t^B) = \delta^{AB}/2$.

For calculating the coupling constant renormalization it is convenient to use the background field method. In the supersymmetric case the background gauge superfield $\bm{V}$, such that $e^{2\bm{V}} = e^{\bm{\Omega}^+} e^{\bm{\Omega}}$, is introduced by the substitution $e^{2V} \to e^{\bm{\Omega}^+} e^{2V} e^{\bm{\Omega}}$.

We regularize the theory (\ref{SYM_Action_Original}) by the BRST invariant version of the higher covariant derivative regularization following Ref. \cite{Aleshin:2016yvj}. In particular, we add to the action (\ref{SYM_Action_Original}) terms with the higher degrees of covariant derivatives, so that

\begin{eqnarray}
&& S+S_{\Lambda} = \frac{1}{2e_0^2}\,\mbox{Re}\,\mbox{tr}\int d^4x\,
d^2\theta\, e^{\Omega} e^{\bm{\Omega}} W^a e^{-\bm{\Omega}}
e^{-\Omega} R\Big(-\frac{\bar\nabla^2
\nabla^2}{16\Lambda^2}\Big)_{Adj} e^{\Omega} e^{\bm{\Omega}}
W_a
e^{-\bm{\Omega}} e^{-\Omega}\qquad\nonumber\\
&& + \frac{1}{4} \int d^4x\, d^4\theta\,\phi^+ e^{\bm{\Omega}^+}
e^{\Omega^+} F\Big(-\frac{\bar\nabla^2
\nabla^2}{16\Lambda^2}\Big) e^{\Omega} e^{\bm{\Omega}}
\phi  + \Big(\frac{1}{6} \int d^4x\,d^2\theta\,\lambda_0^{ijk} \phi_i
\phi_j \phi_k + \mbox{c.c.}\Big),
\end{eqnarray}

\noindent
where the supersymmetric and gauge covariant derivatives are defined by

\begin{equation}
\nabla_a = e^{-\Omega^+} e^{-\bm{\Omega}^+} D_a e^{\bm{\Omega}^+}
e^{\Omega^+}; \qquad \bar\nabla_{\dot a} = e^{\Omega}
e^{\bm{\Omega}} \bar D_{\dot a} e^{-\bm{\Omega}} e^{-\Omega}
\end{equation}

\noindent
with $e^{2V} = e^{\Omega^+} e^\Omega$. The regulator functions $R(x)$ and $F(x)$ should have sufficiently rapid growth at infinity and satisfy the conditions $R(0)=1$ and $F(0)=1$. The gauge fixing term invariant under the background gauge transformations has the form

\begin{equation}
S_{\mbox{\scriptsize gf}} = -\frac{1}{16 \xi_0 e_0^2}\mbox{tr} \int
d^4x\,d^4\theta\,\bm{\nabla}^2 V K\Big(-\frac{\bm{\bar\nabla}^2
\bm{\nabla}^2}{16\Lambda^2}\Big)_{Adj}\bm{\bar \nabla}^2 V,
\end{equation}

\noindent
where $\xi_0$ is the bare gauge parameter, and the background covariant derivatives are given by

\begin{equation}
\bm{\nabla}_a = e^{-\bm{\Omega}^+} D_a e^{\bm{\Omega}^+}; \qquad \bm{\bar\nabla}_{\dot a} = e^{\bm{\Omega}} \bar D_{\dot a} e^{-\bm{\Omega}}.
\end{equation}

\noindent
The regulator $K(x)$ also satisfies the condition $K(0)=1$ and should have sufficiently rapid growth at infinity.

Also it is necessary to introduce the Faddeev--Popov and Niesen--Kallosh ghosts and the Pauli--Villars determinants for regularizing one-loop divergences, which remain after adding the higher derivative terms. The details of these constructions can be found in \cite{Aleshin:2016yvj}. The quantum corrections considered in this paper do not involve these fields, so that we will not discuss them in details. We only note that the actions for the Pauli--Villars superfields are quadratic in the chiral matter superfields. This implies that there are no Yukawa interaction terms including the Pauli--Villars superfields.

Having in mind the exact results derived with the higher derivative regularization for Abelian supersymmetric theories, it is natural to suggest that the NSVZ relation in the non-Abelian case is satisfied by the RG functions defined in terms of the bare couplings if the theory is regularized by higher covariant derivatives. According to \cite{Stepanyantz:2016gtk}, the NSVZ equation can be rewritten in the form of the relation (\ref{New_NSVZ}) between the $\beta$-function and the anomalous dimensions of the quantum gauge superfield, of the Faddeev--Popov ghosts, and of the matter superfields. Eq. (\ref{New_NSVZ}) implies existence of the relation between the Green functions of these superfields, which can be written as

\begin{eqnarray}\label{Green_Function_Relation}
&& \frac{d}{d\ln\Lambda}\Big(d^{-1} - \alpha_0^{-1}\Big)\Big|_{\alpha,\lambda=\mbox{\scriptsize const};\ p\to 0}
= - \frac{3 C_2 - T(R)}{2\pi}\nonumber\\
&&\qquad\quad - \frac{1}{2\pi}\frac{d}{d\ln\Lambda}\Big(- 2C_2 \ln G_c - C_2 \ln G_V + C(R)_i{}^j (\ln G_\phi)_j{}^i/r\Big)\Big|_{\alpha,\lambda=\mbox{\scriptsize const}; q\to 0}.\qquad
\end{eqnarray}

\noindent
This equation admits a simple graphical interpretation \cite{Stepanyantz:2016gtk}. Namely, let us consider a supergraph without external lines. If we attach to it two external lines of the background gauge superfield, then the sum of the diagrams obtained in this way contributes to the function $d^{-1}-\alpha_0^{-1}$. From the other side, various possible cuts of the original supergraph propagators give a set of diagrams contributing to the two-point functions of the quantum gauge superfields, of the Faddeev--Popov ghosts, and of the matter superfields that is to $G_V$, $G_c$, and $(G_\phi)_i{}^j$, respectively. Eq. (\ref{Green_Function_Relation}) relates them to the above described contribution to the function $d^{-1}-\alpha_0^{-1}$.

In this paper we verify that Eq. (\ref{Green_Function_Relation}) is valid for terms proportional to $\lambda_0^4$. Such terms are present in the functions $d^{-1}$ and $(G_\phi)_i{}^j$, which are related to the two-point Green functions of the background gauge superfield and of the matter superfields, respectively. Namely,

\begin{eqnarray}\label{Two_Point_Function}
&& \Gamma^{(2)} - S^{(2)}_{\mbox{\scriptsize gf}} = \frac{1}{4} \int
\frac{d^4p}{(2\pi)^4}\, d^4\theta\, \phi^{*i}(\theta,-p)
\phi_j(\theta,p) G_\phi(\alpha_0,\lambda_0,\Lambda/p)_i{}^j \qquad\nonumber\\
&&\qquad\qquad\qquad - \frac{1}{8\pi}\mbox{tr} \int
\frac{d^4p}{(2\pi)^4}\,d^4\theta\, \bm{V}(\theta,-p)\,\partial^2\Pi_{1/2}
\bm{V}(\theta,p)\,
d^{-1}(\alpha_0,\lambda_0,\Lambda/p)+\ldots.\qquad
\end{eqnarray}

\noindent
The functions $G_c$ and $G_V$ are related to the Green functions of the Faddeev--Popov ghosts and of the quantum gauge superfield. Their definitions are given in \cite{Stepanyantz:2016gtk}, but in this paper these functions are not essential, because they do not contain terms of the considered structure.

If Eq. (\ref{Green_Function_Relation}) is valid, then the NSVZ scheme is given by the prescription (\ref{Scheme_Prescription}). Therefore, we will also be able to verify Eq. (\ref{Scheme_Prescription}) for the considered terms. Note that this check is non-trivial, because we consider the scheme-dependent contributions to the NSVZ relation.

\section{Terms quartic in Yukawa couplings in the NSVZ relation}
\hspace*{\parindent}\label{Section_Calculation}

\begin{figure}[h]
\begin{picture}(0,2)
\put(4,0){\includegraphics[scale=0.08]{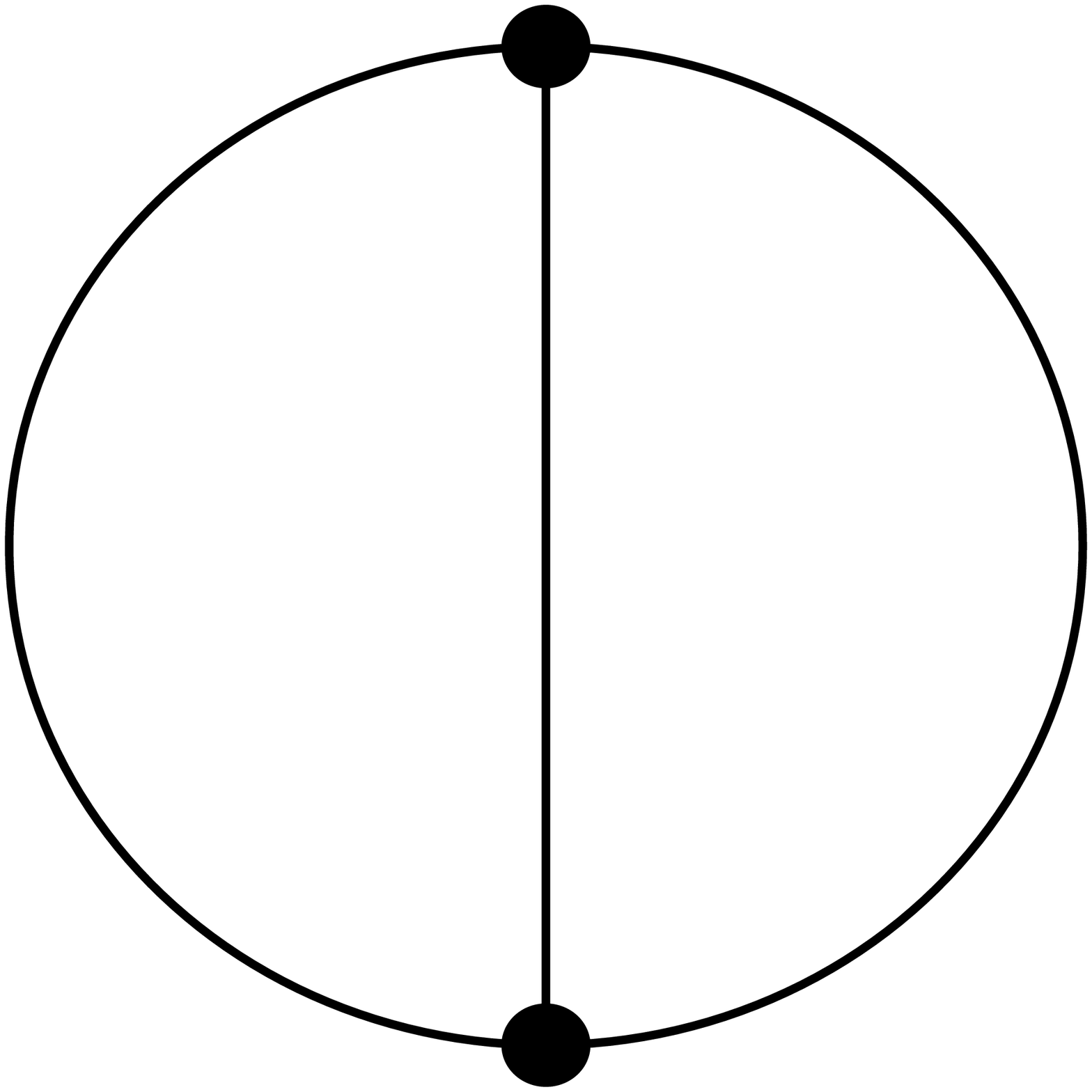}}
\put(7.5,0){\includegraphics[scale=0.08]{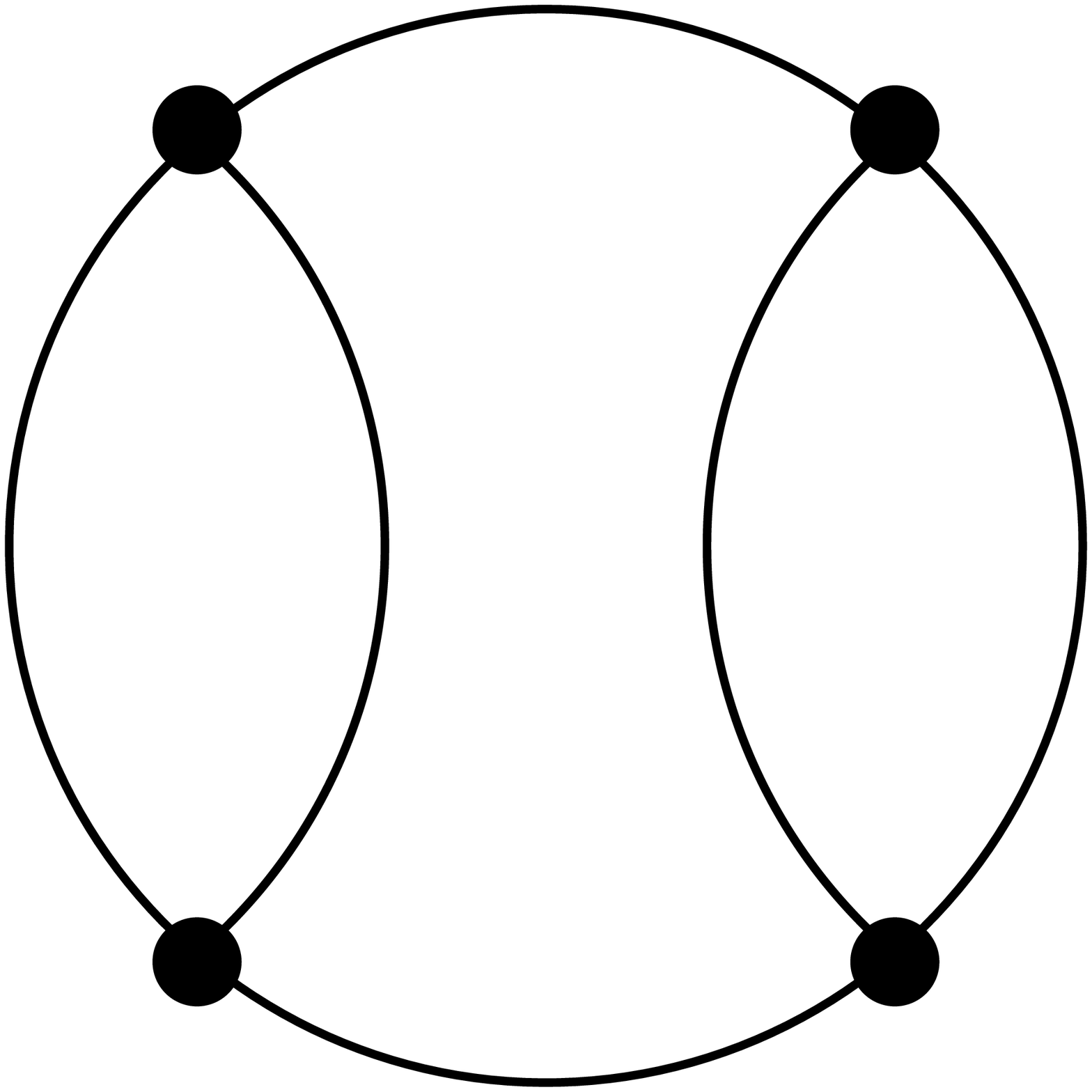}} \put(11,0){\includegraphics[scale=0.08]{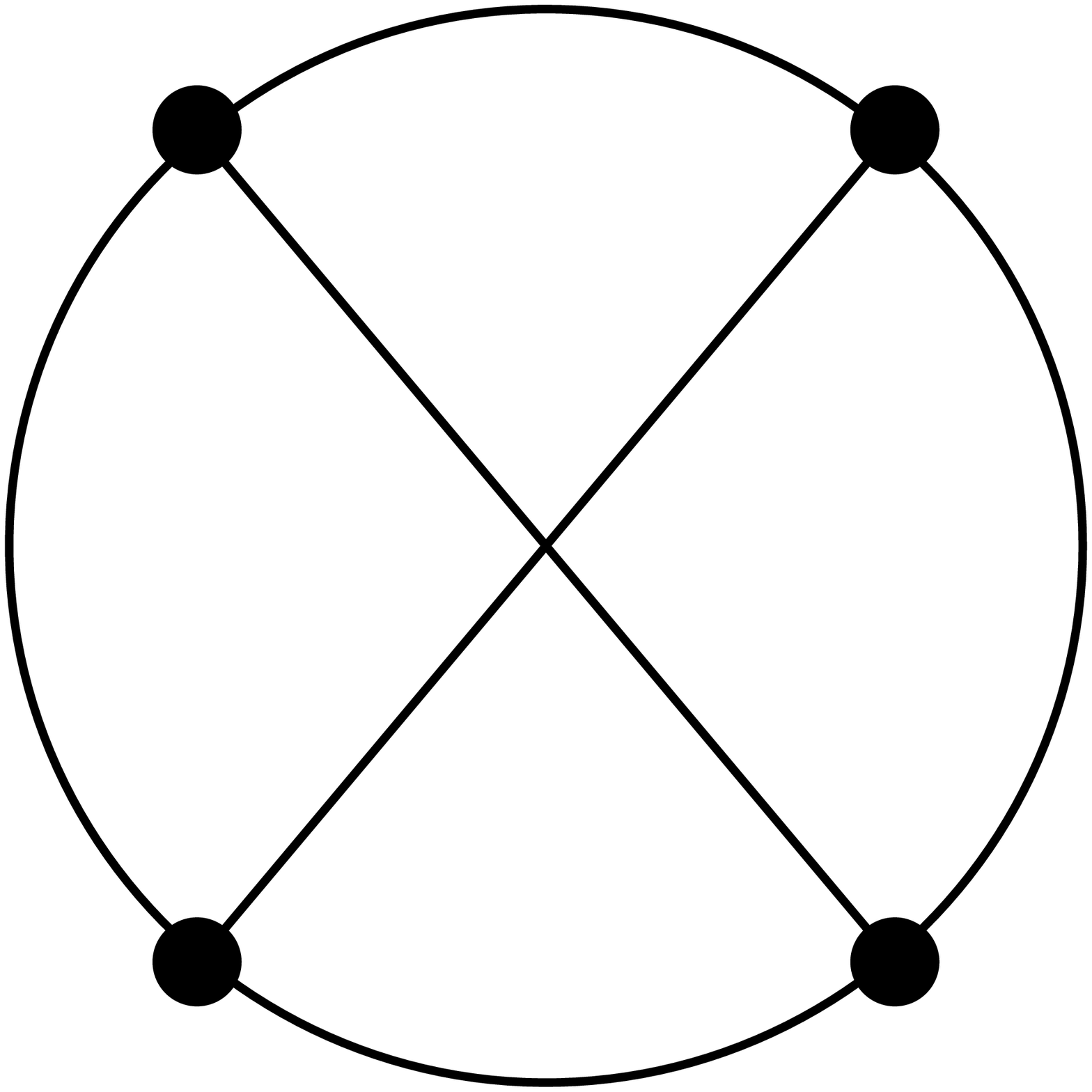}}
\end{picture}
\caption{We consider diagrams which are obtained from the first two graphs by attaching two external lines of the background gauge superfield. The last graph vanishes in the massless case.}\label{Figure_Two_Graphs}
\end{figure}

\begin{figure}[h]
\begin{picture}(0,2)
\put(5.5,0){\includegraphics[scale=0.08]{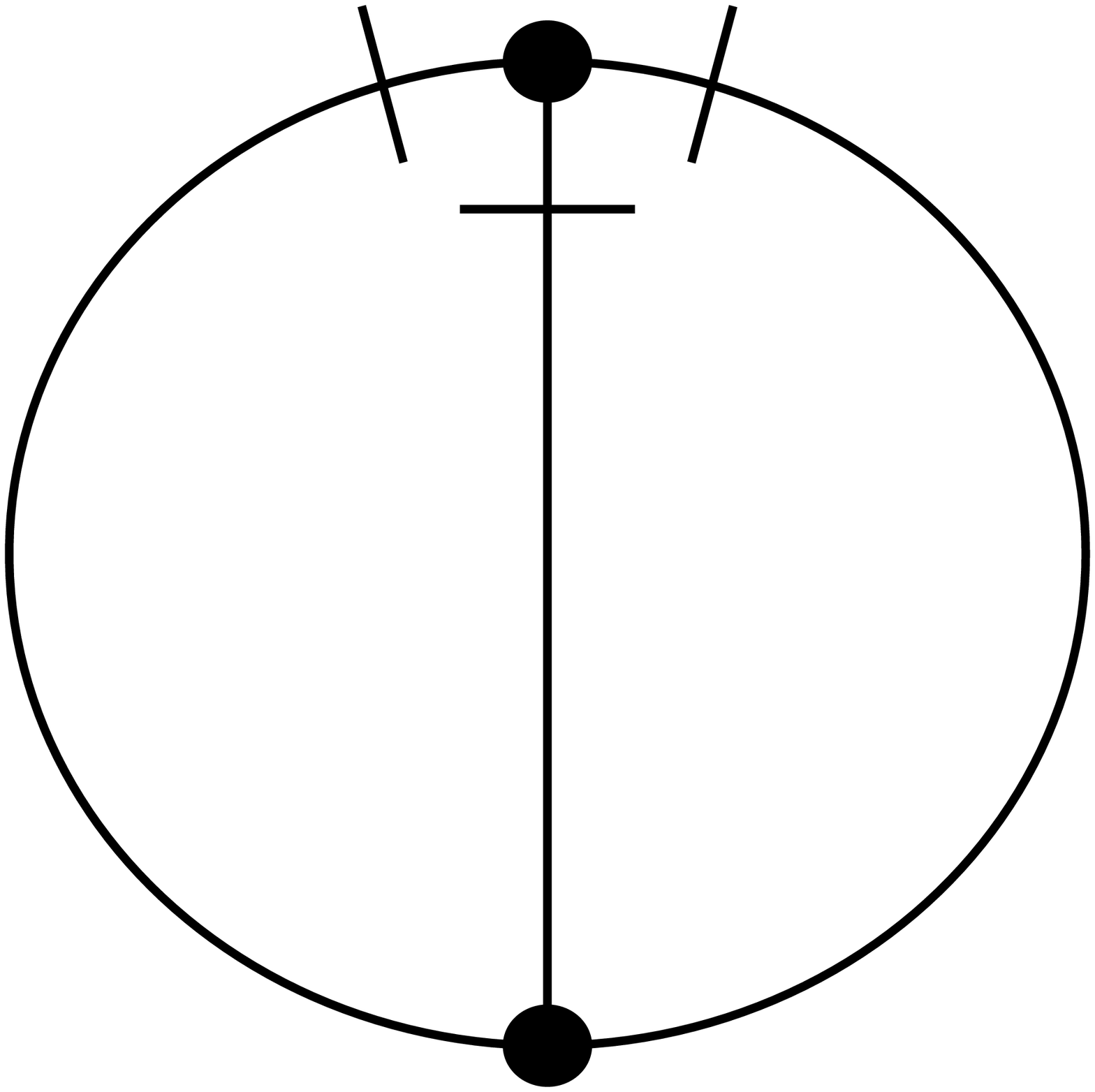}}
\put(9.5,0){\includegraphics[scale=0.08]{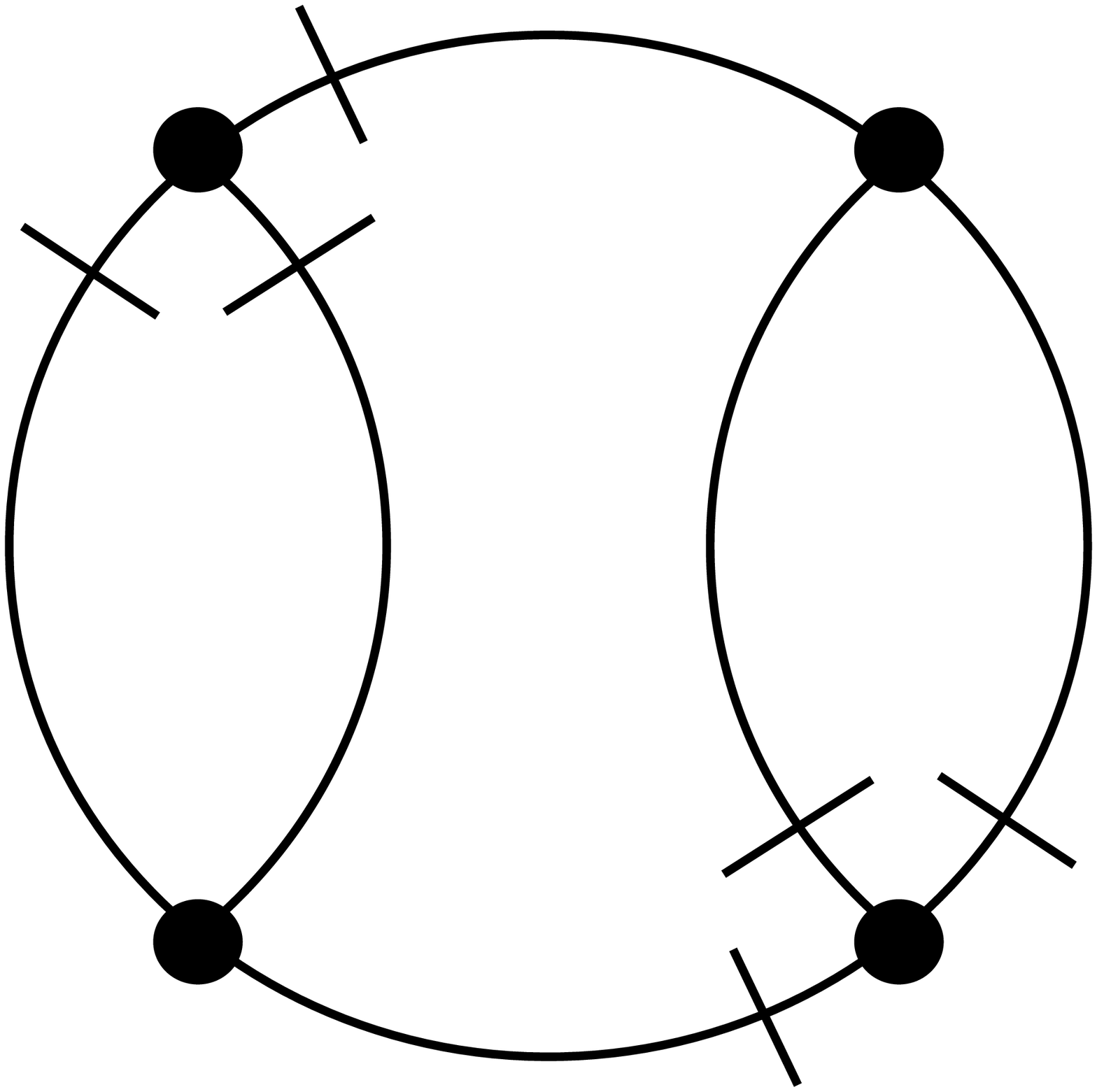}}
\end{picture}
\caption{Here we marked chiral ends of propagators for the graphs which do not vanish in the massless case.}\label{Figure_Graph}
\end{figure}

In this paper we are interested in terms quartic in the Yukawa couplings (without the gauge coupling constant) in the NSVZ relation (\ref{New_NSVZ}). Below we will see that  for calculating them, it is also necessary to know terms quadratic in the Yukawa couplings without the gauge coupling constant. All terms mentioned above
correspond to one two-loop graph and two three-loop graphs presented in Fig. \ref{Figure_Two_Graphs}. However, in the massless limit the last graph vanishes. Really, in the massless theory each propagator has a chiral end and an antichiral end. Each vertex connects either three chiral ends or three antichiral ends of the propagators. However, one can easily see that it is impossible to satisfy both these requirements in the last graph. The other graphs nontrivially contribute in the massless case. The arrangement of chiral and antichiral vertices for these graphs is presented in Fig. \ref{Figure_Graph}.

\begin{figure}[h]
\begin{picture}(0,2.5)
\put(3.0,0.3){\includegraphics[scale=0.15]{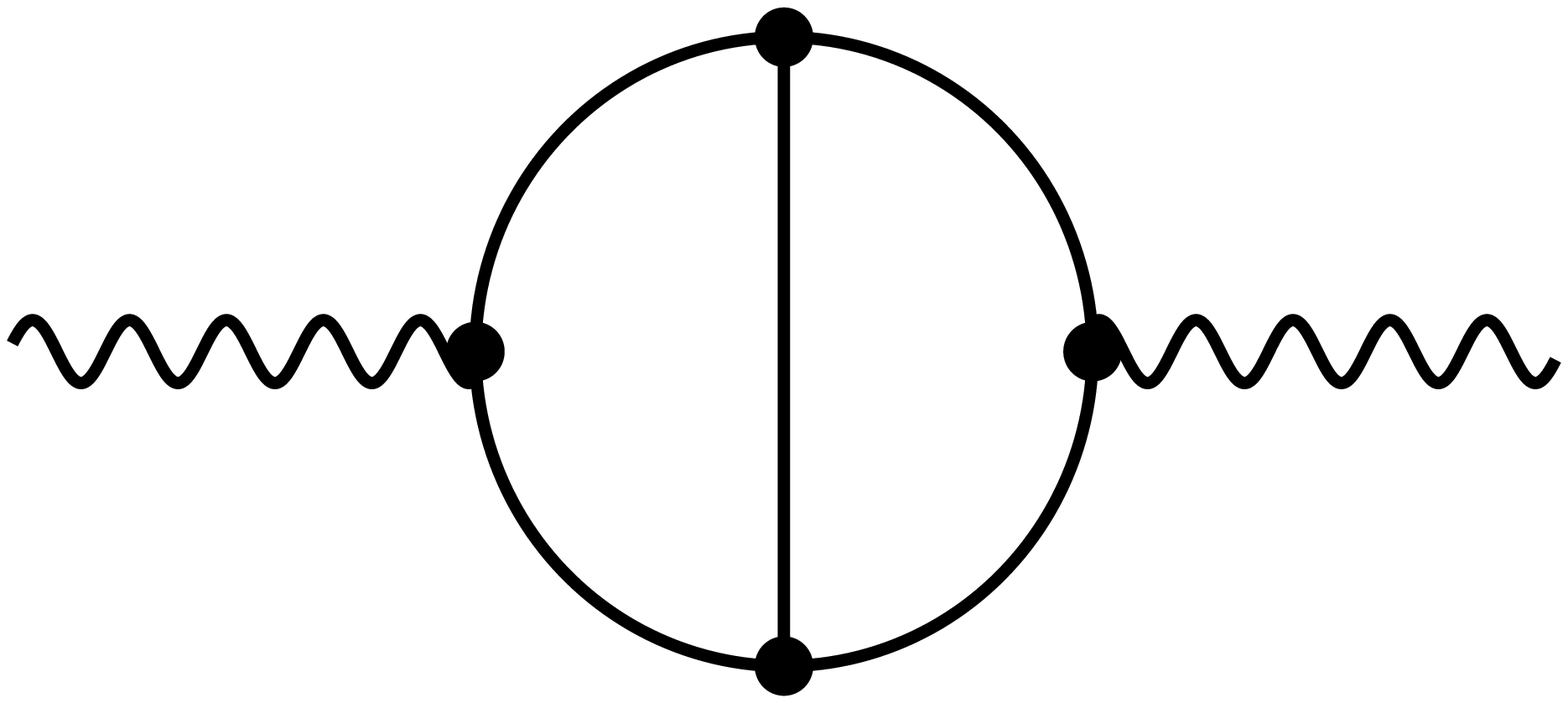}}
\put(3.0,1.9){$(1)$}
\put(7.0,0.3){\includegraphics[scale=0.15]{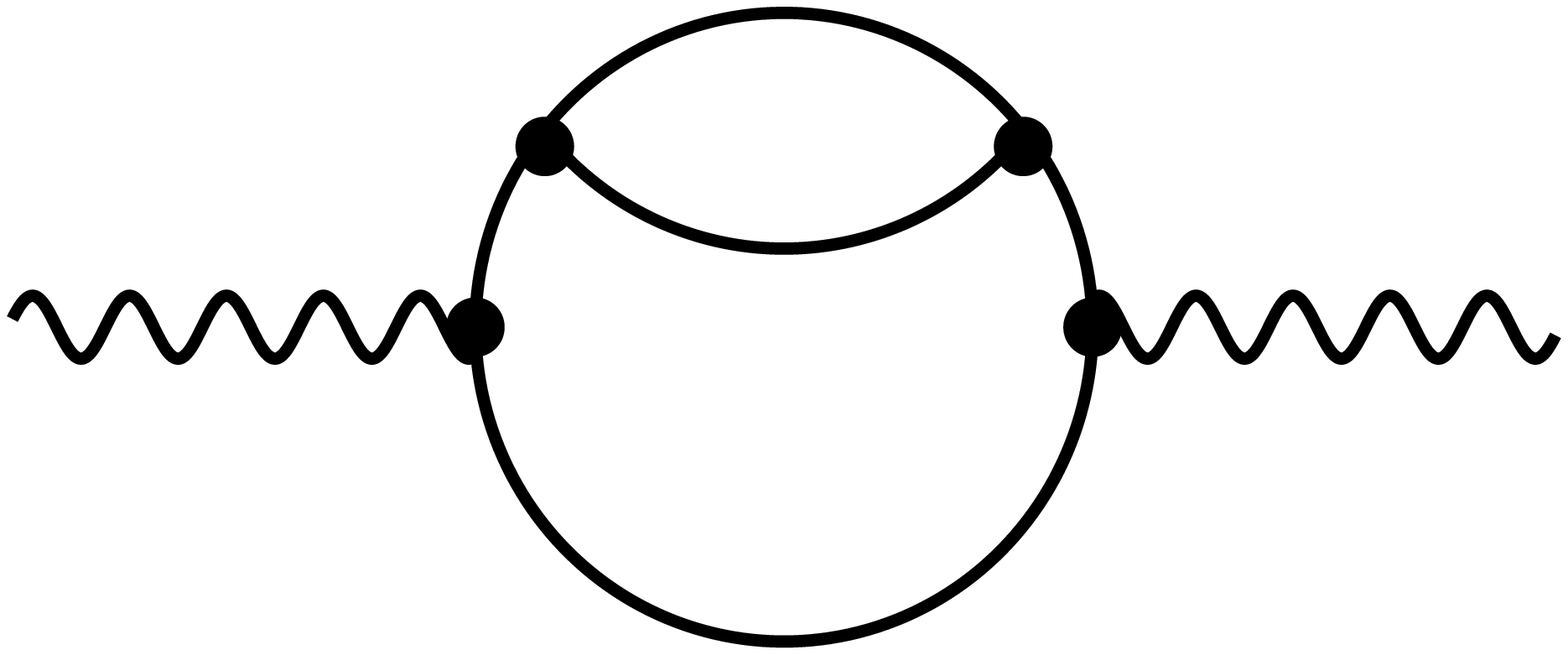}}
\put(7.0,1.9){$(2)$}
\put(11,0){\includegraphics[scale=0.15]{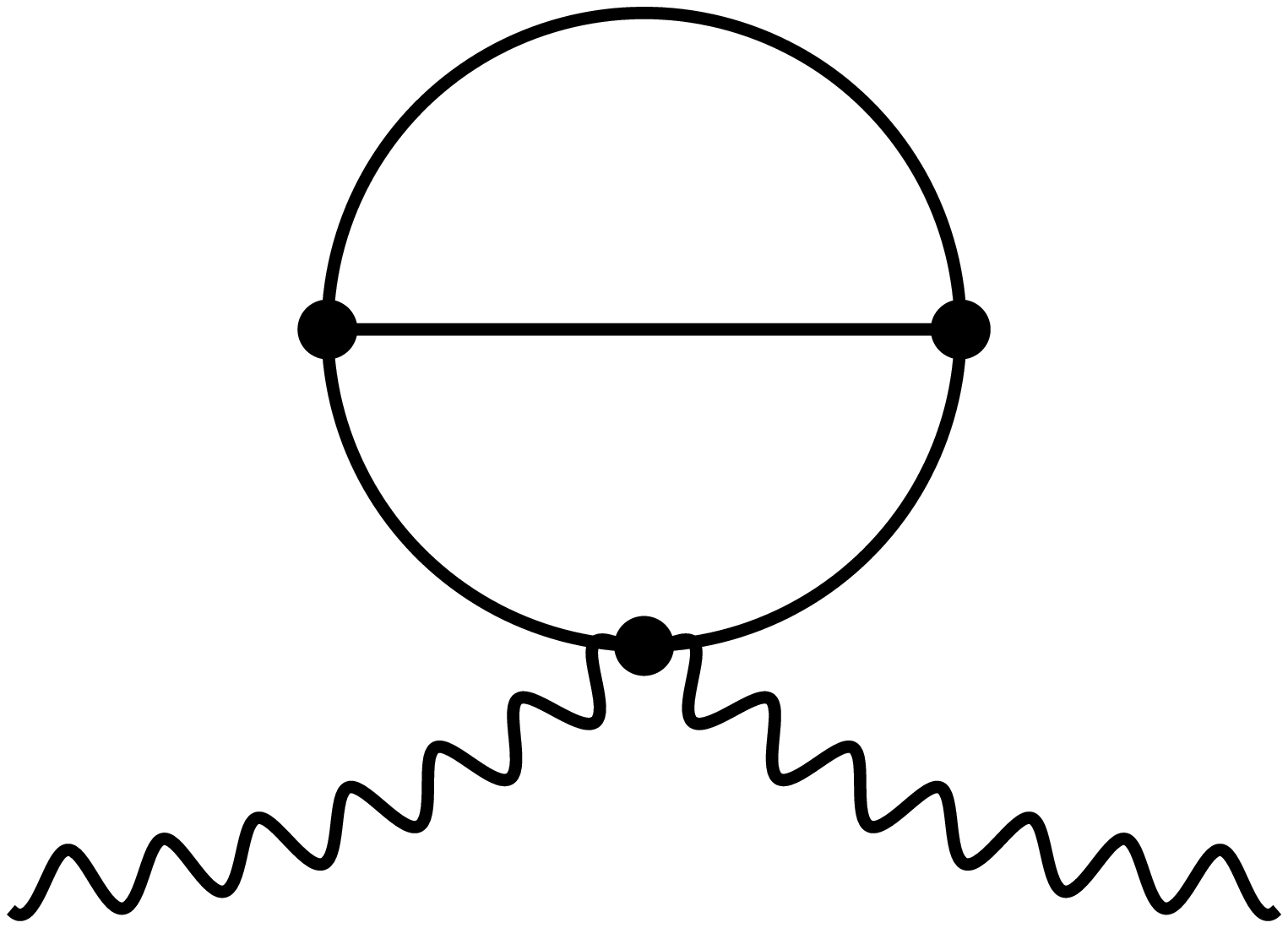}}
\put(11,1.9){$(3)$}
\end{picture}
\caption{Diagrams giving the two-loop contribution quadratic in the Yukawa couplings to the $\beta$-function. The wavy lines correspond to the background gauge superfield $\bm{V}$.}\label{Figure_Beta_Two_Loop}
\end{figure}

\begin{figure}[h]
\begin{picture}(0,5)
\put(0.5,2.1){\includegraphics[scale=0.15]{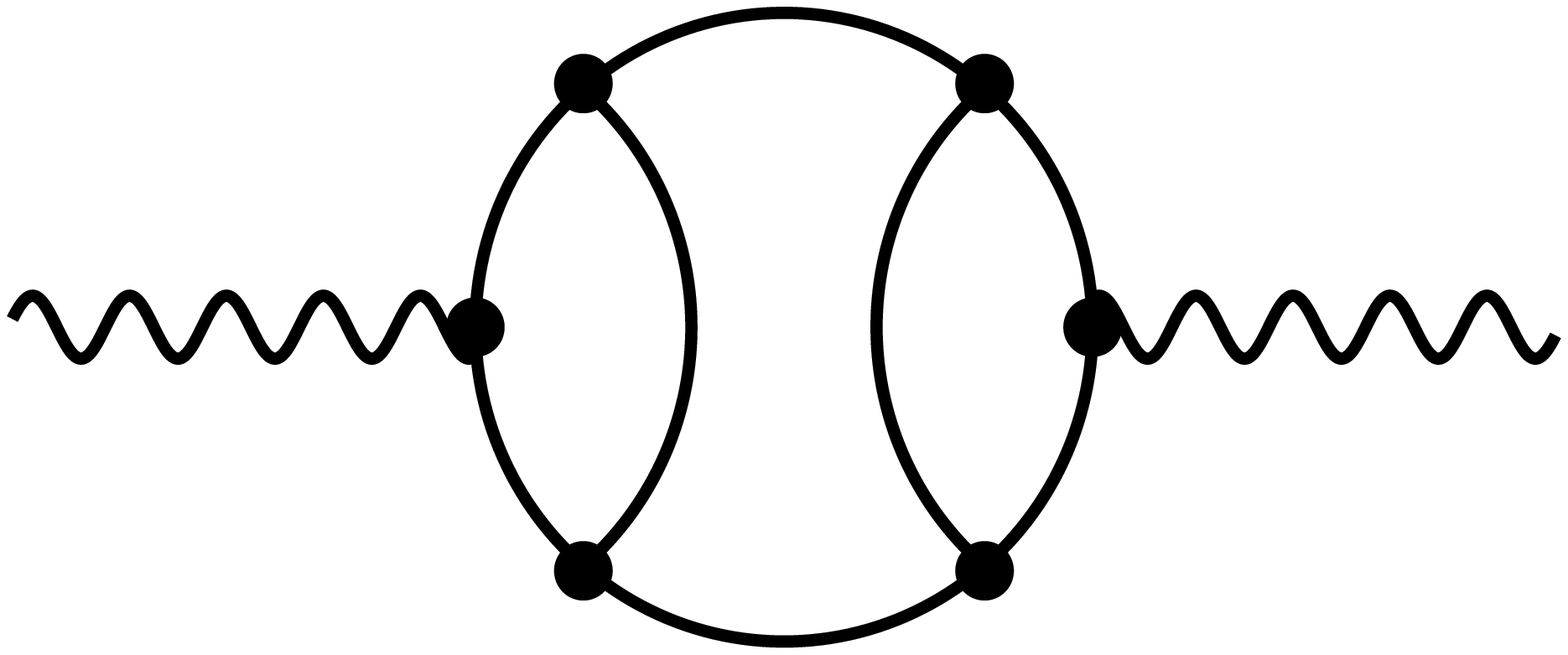}}
\put(0.5,4.4){$(1)$}
\put(4.5,2.1){\includegraphics[scale=0.15]{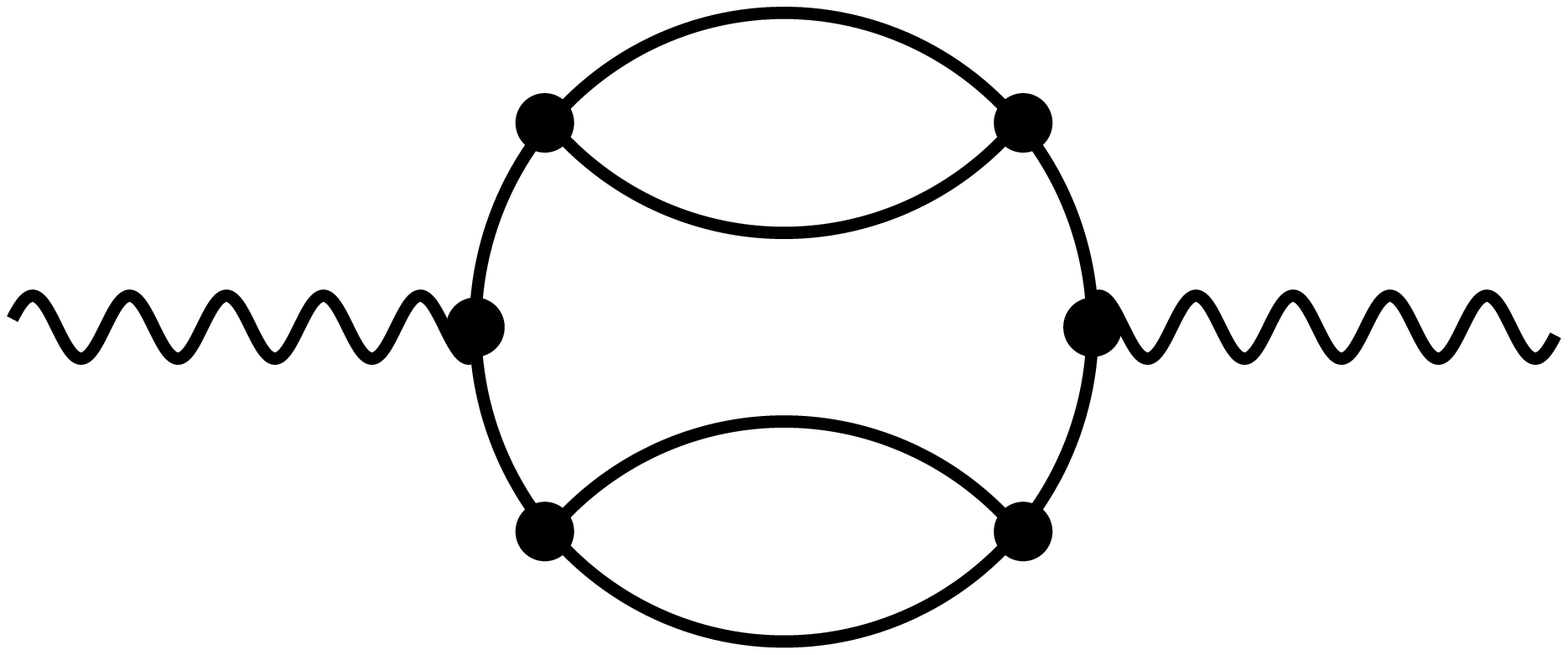}}
\put(4.5,4.4){$(2)$}
\put(8.5,2.1){\includegraphics[scale=0.15]{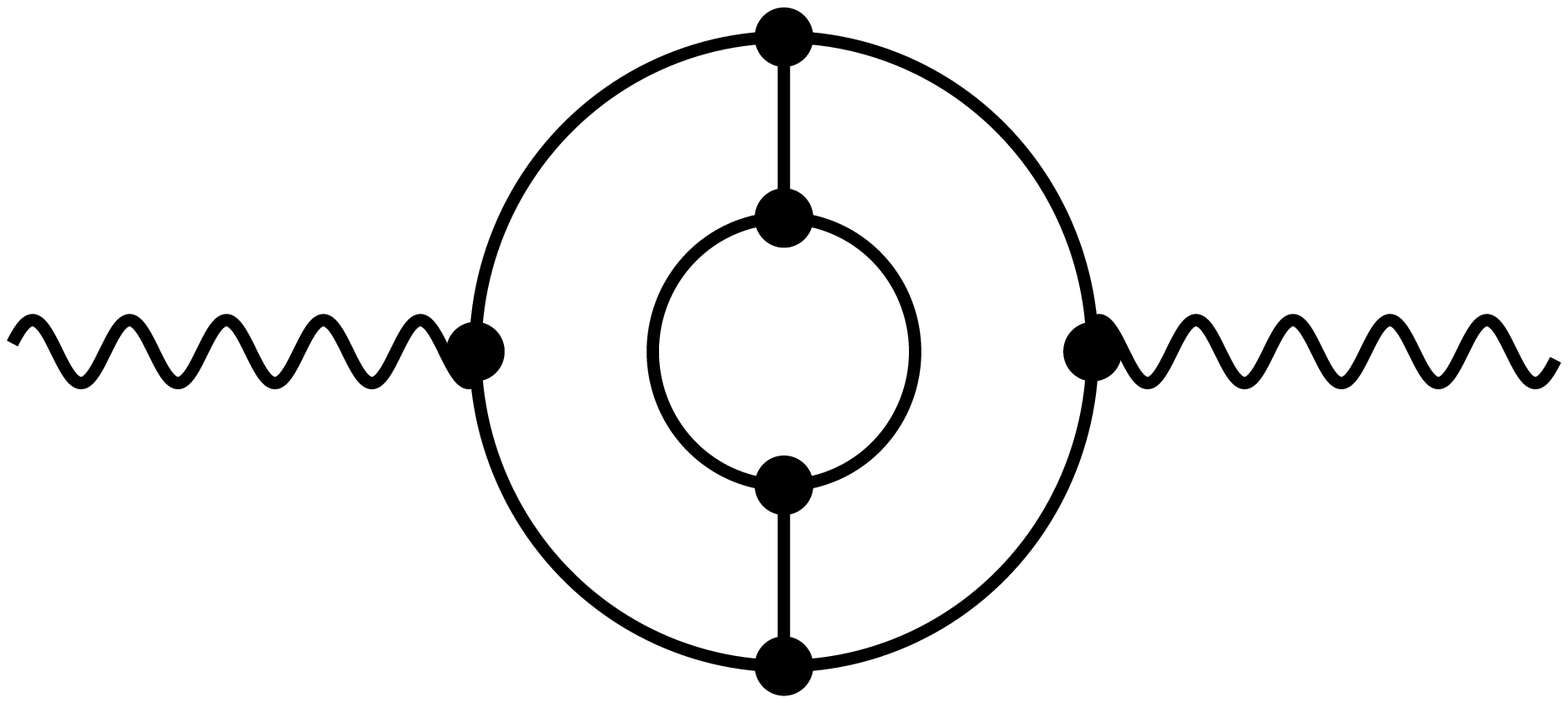}}
\put(8.5,4.4){$(3)$}
\put(12.5,2.1){\includegraphics[scale=0.15]{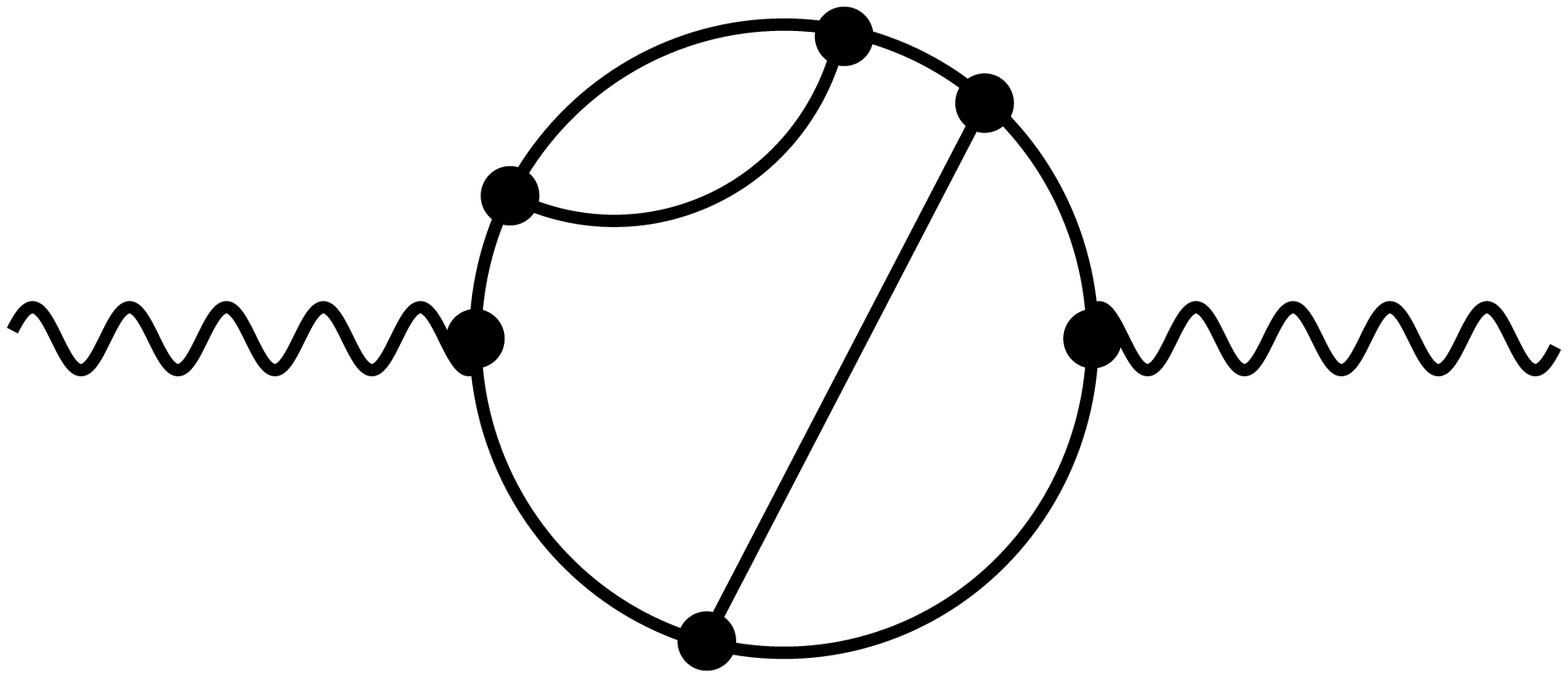}}
\put(12.5,4.4){$(4)$}
\put(0.5,0){\includegraphics[scale=0.15]{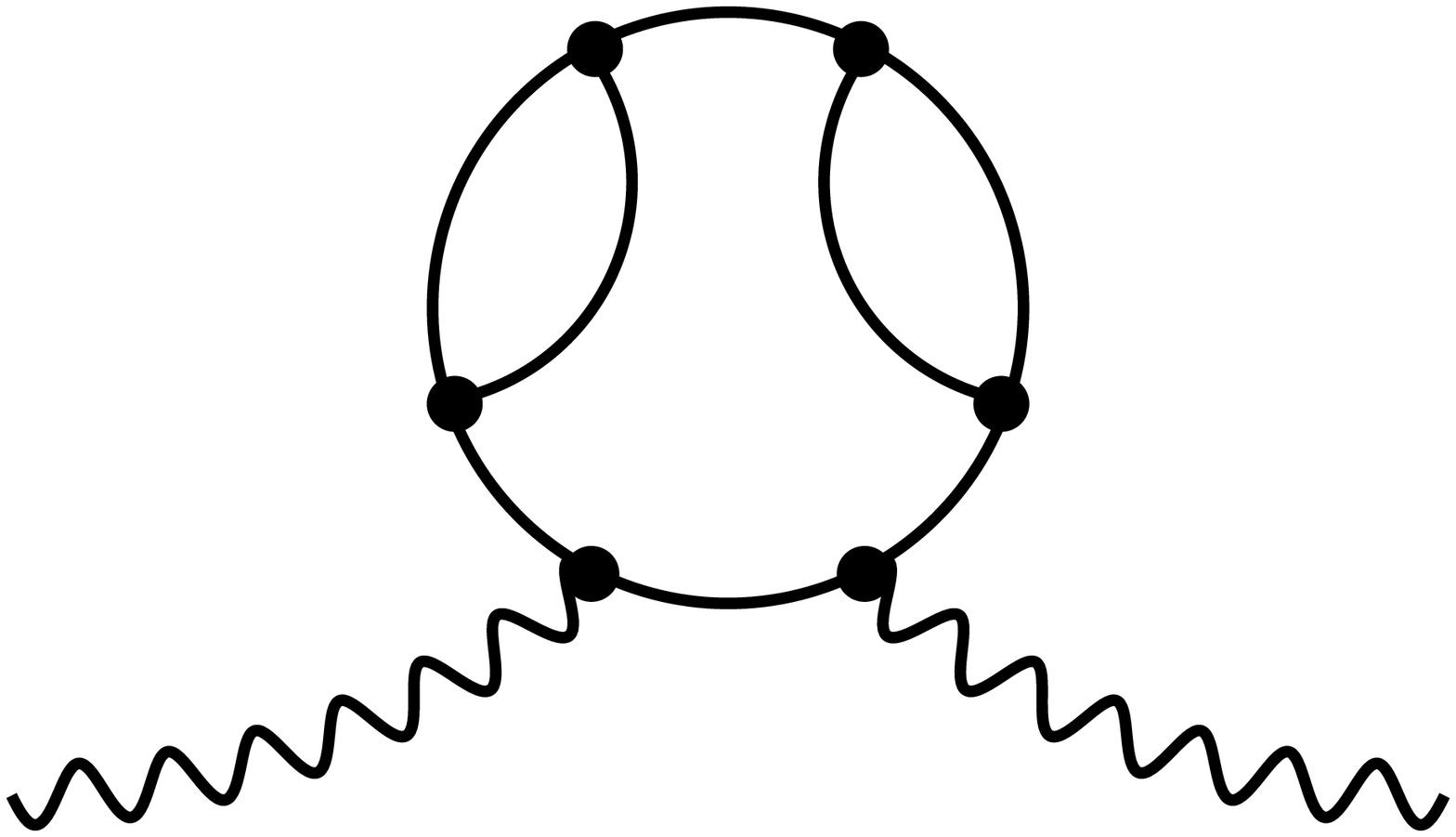}}
\put(0.5,1.9){$(5)$}
\put(4.5,0){\includegraphics[scale=0.15]{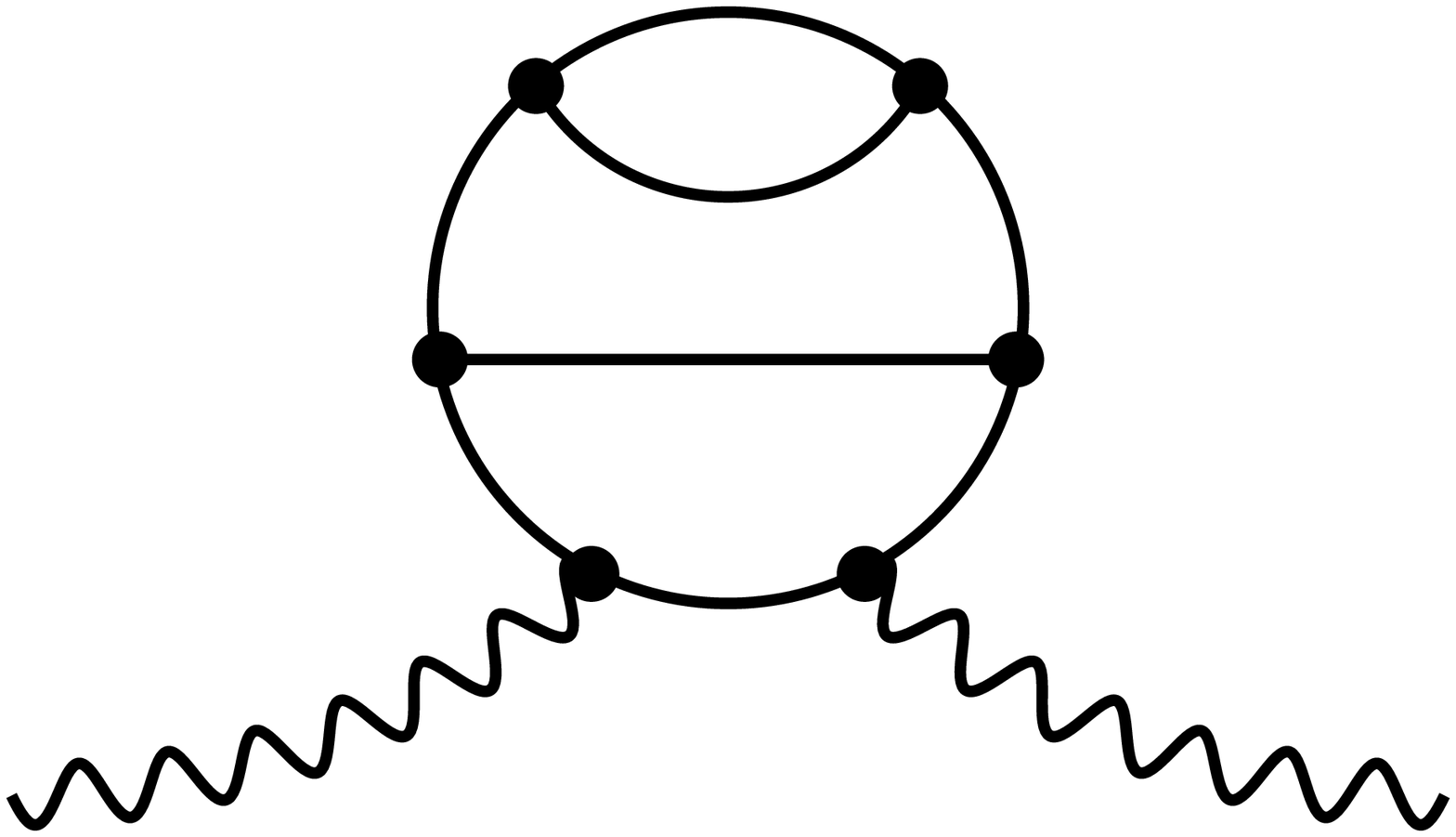}}
\put(4.5,1.9){$(6)$}
\put(8.5,0){\includegraphics[scale=0.15]{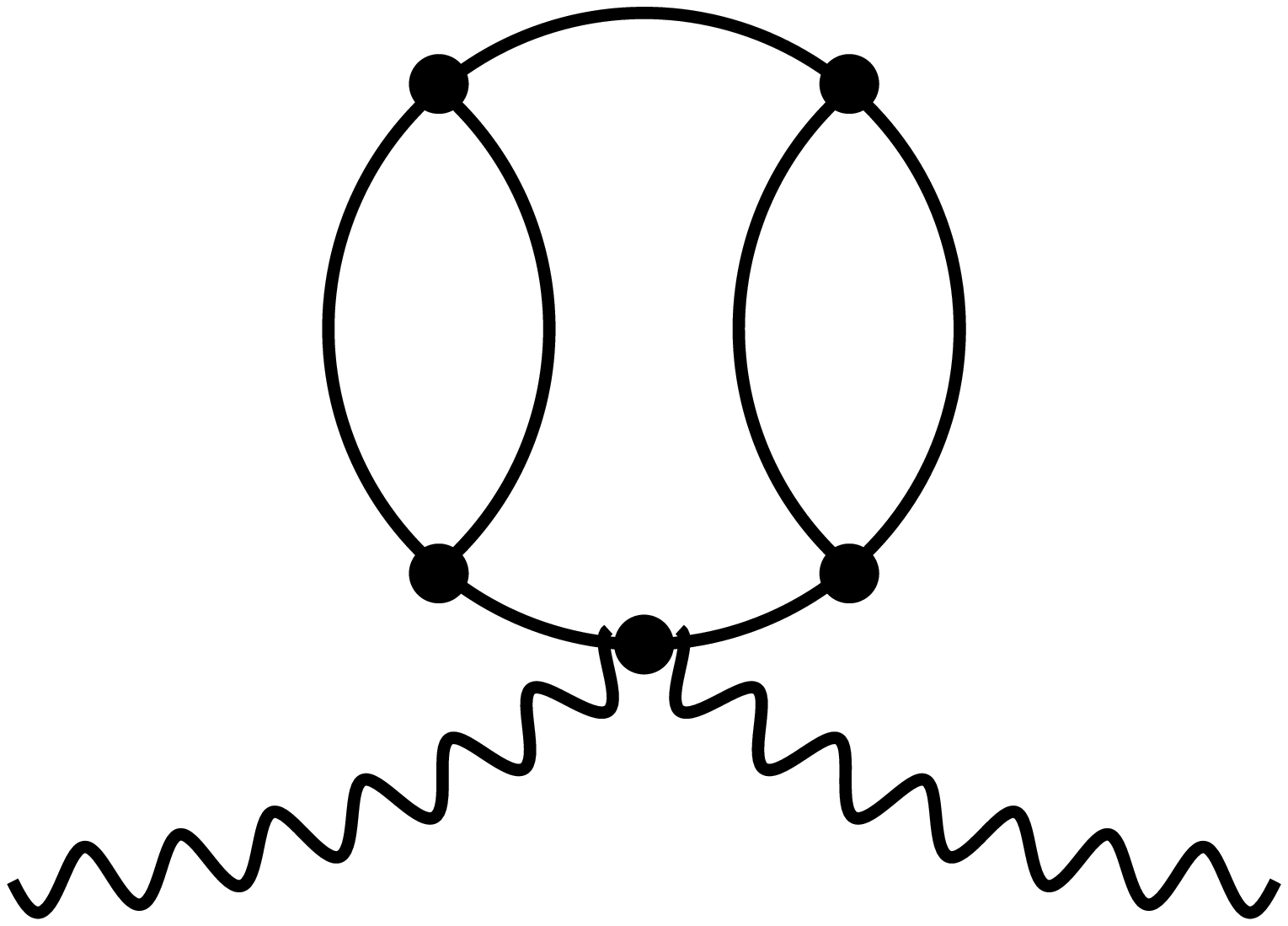}}
\put(8.5,1.9){$(7)$}
\put(12.5,0){\includegraphics[scale=0.15]{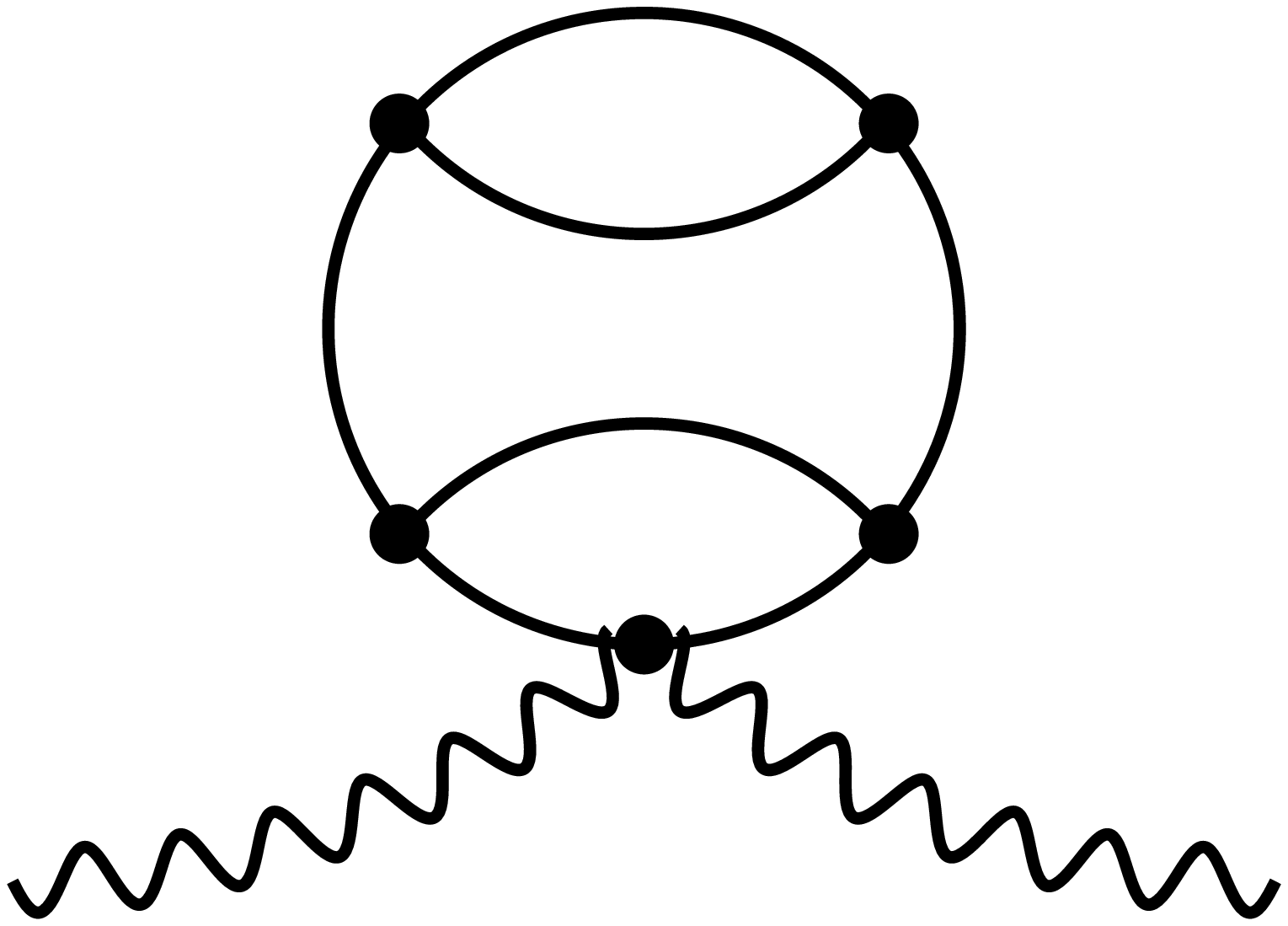}}
\put(12.5,1.9){$(8)$}
\end{picture}
\caption{These diagrams give the three-loop contribution quartic in the Yukawa couplings to the $\beta$-function.}\label{Figure_Beta}
\end{figure}

\begin{figure}[h]
\begin{picture}(0,2)
\put(4,-0.8){\includegraphics[scale=0.16]{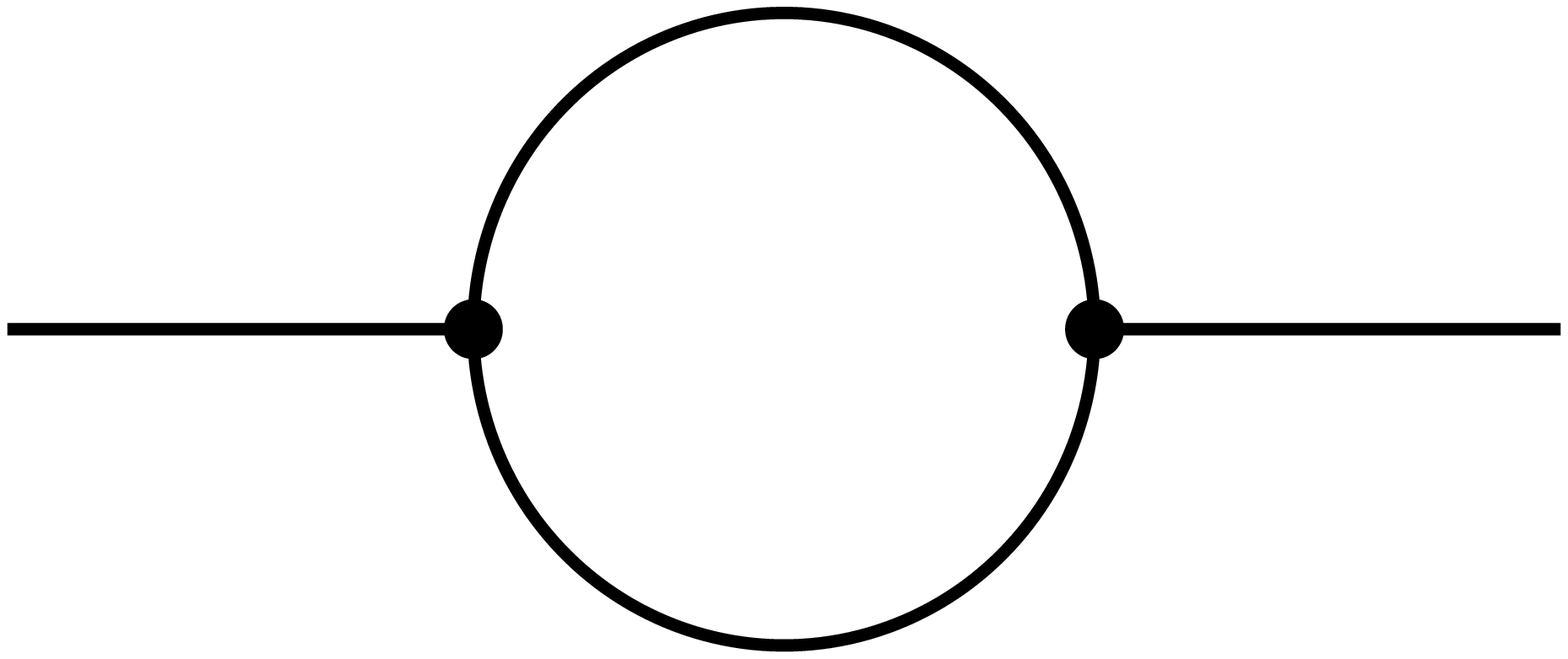}}
\put(9,-0.8){\includegraphics[scale=0.16]{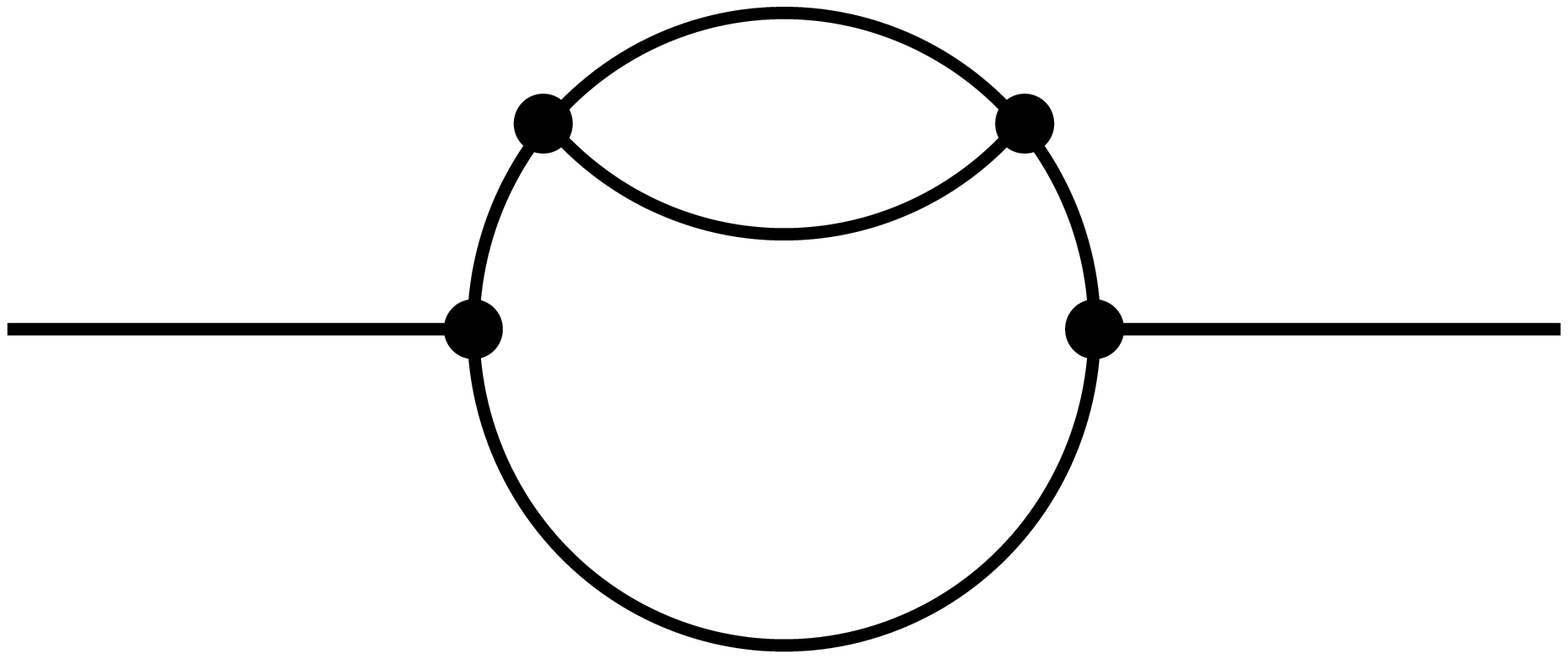}}
\end{picture}
\caption{These diagrams give the one and two-loop contributions to the anomalous dimension of the matter superfields quadratic and quartic in Yukawa couplings, respectively.}\label{Figure_Gamma}
\end{figure}

\begin{figure}[h]
\begin{picture}(0,2)
\put(6.5,0.0){\includegraphics[scale=0.16]{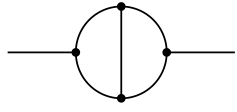}}
\end{picture}
\caption{This superdiagram is obtained after cutting a matter line in the third graph in Fig. \ref{Figure_Two_Graphs}. It is easy to see that it gives vanishing contribution to the anomalous dimension in the massless case.}\label{Figure_Vanishing_Gamma}
\end{figure}

As we have explained above, to obtain the diagrams contributing to the $\beta$-function from the graphs presented in Fig. \ref{Figure_Graph}, it is necessary to attach two external lines of the background gauge superfield $\bm{V}$ by all possible ways. This gives three two-loop diagrams presented in Fig. \ref{Figure_Beta_Two_Loop} and eight three-loop diagrams presented in Fig. \ref{Figure_Beta}. Their contribution should be compared with the part of the anomalous dimension of the matter superfield which comes from the diagrams obtained by all possible cuts of the graphs presented in Fig. \ref{Figure_Graph}. Certainly, it is necessary to take into account only the 1PI graphs, which are presented in Fig. \ref{Figure_Gamma}, because the effective action encodes the sum of 1PI graphs. Note that cutting a matter line in the (vanishing in the massless limit) third graph in Fig. \ref{Figure_Two_Graphs} gives the only superdiagram presented in Fig. \ref{Figure_Vanishing_Gamma}. One can easily check that in the massless limit it vanishes and, therefore, does not contribute to the anomalous dimension.

Let us start with calculating the diagrams presented in Figs. \ref{Figure_Beta_Two_Loop} and \ref{Figure_Beta}. More exactly, we will calculate their contribution to the $\beta$-function defined in terms of the bare coupling constant,

\begin{equation}\label{D_Derivative}
\frac{d}{d\ln\Lambda} \Big(d^{-1}(\alpha_0,\lambda_0,\Lambda/p)-\alpha_0^{-1}\Big)\Big|_{p=0} = \frac{\beta(\alpha_0,\lambda_0)}{\alpha_0^2}.
\end{equation}

\noindent
The differentiation with respect to $\ln\Lambda$ in this expression should be made at fixed values of the renormalized gauge and Yukawa couplings, while the result should be reexpressed in terms of the bare ones. Note that it is also necessary to take the limit $p\to 0$, where $p$ is the external momentum, in order to get rid of the finite terms proportional to $\Lambda^{-k}$, where $k$ is a positive integer.

The results for contributions of all diagrams presented in Figs. \ref{Figure_Beta_Two_Loop} and \ref{Figure_Beta} to the effective action in the limit of the vanishing external momentum are collected in Appendix \ref{Appendix_Explicit_Graphs}. Their sum appears to be transversal as it should be due to the background gauge invariance. We have also verified that it is given by an integral of a double total derivative. In particular, the contribution of the considered supergraphs to the expression (\ref{D_Derivative}) can be written as\footnote{For simplicity, we do not include the one-loop contribution into this expression.}

\begin{eqnarray}\label{Delta_Beta_Integral}
&&\hspace*{-5mm} \frac{\Delta\beta(\alpha_0,\lambda_0)}{\alpha_0^2} = -\frac{2\pi}{r} C(R)_i{}^j \frac{d}{d\ln\Lambda} \int \frac{d^4k}{(2\pi)^4} \frac{d^4q}{(2\pi)^4} \,\lambda_0^{imn} \lambda^*_{0jmn} \frac{\partial}{\partial q_\mu} \frac{\partial}{\partial q^\mu} \Big(\frac{1}{k^2 F_k\, q^2 F_q\, (q+k)^2 F_{q+k}}\Big)\nonumber\\
&&\hspace*{-5mm} + \frac{4\pi}{r} C(R)_i{}^j \frac{d}{d\ln\Lambda} \int \frac{d^4k}{(2\pi)^4} \frac{d^4l}{(2\pi)^4} \frac{d^4q}{(2\pi)^4} \Bigg(\lambda_0^{iab}\lambda^*_{0kab} \lambda_0^{kcd}\lambda^*_{0jcd} \Big(\frac{\partial}{\partial k_\mu} \frac{\partial}{\partial k^\mu} - \frac{\partial}{\partial q_\mu} \frac{\partial}{\partial q^\mu}\Big)\nonumber\\
&&\hspace*{-5mm} + 2 \lambda_0^{iab}\lambda^*_{0jac} \lambda_0^{cde}\lambda^*_{0bde}\, \frac{\partial}{\partial q_\mu} \frac{\partial}{\partial q^\mu} \Bigg) \frac{1}{k^2 F_k^2\, q^2 F_q\, (q+k)^2 F_{q+k}\, l^2 F_l\, (l+k)^2 F_{l+k}},
\end{eqnarray}

\noindent
where the derivative with respect to $\ln\Lambda$ is calculated at fixed values of the renormalized Yukawa constants.\footnote{Note that Eq. (\ref{Delta_Beta_Integral}) is not contributed by the Pauli--Villars superfields, because, for the considered regularization \cite{Aleshin:2016yvj}, there are no triple vertices which include the Pauli--Villars superfields.} To write the complete $\beta$-function, it is necessary to add the one-loop contribution and the contributions of the other supergraphs, which have not been considered in this paper. The result can be presented in the form

\begin{equation}\label{Total_Beta}
\frac{\beta(\alpha_0,\lambda_0)}{\alpha_0^2} = -\frac{1}{2\pi}\Big(3C_2-T(R)\Big) + \frac{\Delta\beta(\alpha_0,\lambda_0)}{\alpha_0^2} + O(\alpha_0) + O(\lambda_0^6),
\end{equation}

\noindent
where $O(\alpha_0)$ denotes terms proportional to $\alpha_0$ (including the ones which appear in the two-loop approximation) and $O(\lambda_0^6)$ denotes terms with higher degrees of the Yukawa couplings in higher orders. The two-loop part of the result agrees with the expression obtained in \cite{Shevtsova2009,Pimenov:2009hv} for the particular case $F(x) = 1+x^m$ and for a different version of the higher derivative regularization\footnote{For the considered terms the difference of the regularizations is not essential.}, which has been subsequently written as an integral of double total derivative in \cite{Stepanyantz:2011bz}.

The expression (\ref{Delta_Beta_Integral}) does not vanish because of singularities of the integrand. This can be illustrated by a simple example,

\begin{equation}
\int \frac{d^4q}{(2\pi)^4} \frac{\partial}{\partial q^\mu} \Big(\frac{q^\mu}{q^4} f(q^2)\Big) = -\frac{1}{8\pi^2} f(0),
\end{equation}

\noindent
where we assume that the function $f(q^2)$ is non-singular and has a sufficiently rapid fall-off at infinity. Calculating one of the loop integrals in Eq. (\ref{Delta_Beta_Integral}) by the help of similar equations, we obtain the considered part of the $\beta$-function in the form

\begin{eqnarray}\label{Beta_Integral}
&&\hspace*{-6mm} \frac{\Delta\beta(\alpha_0,\lambda_0)}{\alpha_0^2} = \frac{1}{\pi r} C(R)_i{}^j \frac{d}{d\ln\Lambda}\Big[- \,\lambda_0^{imn} \lambda^*_{0jmn} \int \frac{d^4k}{(2\pi)^4}
\frac{1}{k^4 F_k^2} + \lambda_0^{iab}\lambda^*_{0kab} \lambda_0^{kcd}\lambda^*_{0jcd} \int \frac{d^4k}{(2\pi)^4} \nonumber\\
&&\hspace*{-6mm} \times \int \frac{d^4l}{(2\pi)^4} \frac{1}{k^4 F_k^2\, l^4 F_l^2} + 4\,\lambda_0^{iab}\lambda^*_{0jac} \lambda_0^{cde}\lambda^*_{0bde}\,  \int \frac{d^4k}{(2\pi)^4} \frac{d^4l}{(2\pi)^4} \frac{1}{k^4 F_k^3\, l^2 F_l\, (k+l)^2 F_{k+l}}\Big].
\end{eqnarray}

\noindent
Note that this integral is well-defined due to the differentiation with respect to $\ln\Lambda$ which should be made before the integrations. This will be demonstrated below.

Now, let us compare Eq. (\ref{Beta_Integral}) with the corresponding contribution to the anomalous dimension, which comes from the diagrams presented in Fig. \ref{Figure_Gamma}. Calculating them, we obtain

\begin{eqnarray}\label{G_Integral}
&& G_\phi(\alpha_0,\lambda_0,p/\Lambda)_j{}^i = \delta_j^i + \lambda_0^{iab} \lambda^*_{0jab} \int \frac{d^4k}{(2\pi)^4} \frac{2}{k^2 F_k\, (k+p)^2 F_{k+p}} - \lambda_0^{iab} \lambda^*_{0jac} \lambda_0^{cde} \lambda^*_{0bde}\qquad\nonumber\\
&& \times \int \frac{d^4k}{(2\pi)^4} \frac{d^4l}{(2\pi)^4} \frac{8}{k^2 F_k^2\, (k+p)^2 F_{k+p}\, l^2 F_l\, (k+l)^2 F_{k+l}} + O(\alpha_0) + O(\lambda_0^6).\qquad
\end{eqnarray}

\noindent
Taking the logarithm of this expression and making the differentiation with respect to $\ln\Lambda$ in the limit of the vanishing external momentum, we construct the anomalous dimension defined in terms of the bare couplings,

\begin{equation}\label{RG_Gamma_Equation}
\gamma_\phi(\alpha_0,\lambda_0)_j{}^i = - \frac{d(\ln Z_\phi)_j{}^i}{d\ln\Lambda} = \frac{d(\ln G_\phi)_j{}^i}{d\ln\Lambda}\Big|_{p=0}.
\end{equation}

\noindent
From this equation we obtain the considered part of the anomalous dimension in the form of the sum of loop integrals,

\begin{eqnarray}\label{Gamma_Integral}
&&\hspace*{-7mm} \Delta\gamma_\phi(\lambda_0)_j{}^i = \frac{d}{d\ln\Lambda}\Bigg(
\lambda_0^{iab} \lambda^*_{0jab} \int \frac{d^4k}{(2\pi)^4} \frac{2}{k^4 F_k^2}
- \lambda_0^{iab} \lambda^*_{0kab}  \lambda_0^{kcd} \lambda^*_{0jcd} \int \frac{d^4k}{(2\pi)^4} \frac{d^4l}{(2\pi)^4} \frac{2}{k^4 F_k^2\, l^4 F_{l}^2}\nonumber\\
&&\hspace*{-7mm} - \lambda_0^{iab} \lambda^*_{0jac} \lambda_0^{cde} \lambda^*_{0bde} \int \frac{d^4k}{(2\pi)^4} \frac{d^4l}{(2\pi)^4} \frac{8}{k^4 F_k^3\, l^2 F_l\, (k+l)^2 F_{k+l}}\Bigg).\vphantom{\Bigg(}\qquad
\end{eqnarray}

\noindent
The complete expression for the anomalous dimension also contains terms proportional to $\alpha_0$ (starting from the one-loop approximation) and terms, proportional to $\lambda_0^6$ (starting from the three-loop approximation),

\begin{equation}\label{Complete_Gamma}
\gamma_\phi(\alpha_0,\lambda_0)_i{}^j = \Delta\gamma_\phi(\lambda_0)_i{}^j + O(\alpha_0) + O(\lambda_0^6).
\end{equation}

The expression (\ref{Gamma_Integral}) should be compared with Eq. (\ref{Beta_Integral}). Exactly as in Eq. (\ref{Beta_Integral}), the derivative with respect to $\ln\Lambda$ should be calculated at fixed values of the renormalized Yukawa couplings $\lambda$. Moreover, it is easy to see that the integrals coincide up to the multiplicative factor,

\begin{equation}\label{Three-Loop_NSVZ}
\frac{\Delta\beta(\alpha_0,\lambda_0)}{\alpha_0^2} = -\frac{1}{2\pi r} C(R)_i{}^j \Delta\gamma_\phi(\lambda_0)_j{}^i.
\end{equation}

\noindent
This implies that the NSVZ relation (\ref{New_NSVZ}) (and, therefore, Eq. (\ref{NSVZ})) is satisfied by the RG functions defined in terms of the bare coupling constant for the considered groups of diagrams in the case of using the higher covariant derivative regularization.

\section{Explicit expression for the considered part of the anomalous dimension}
\hspace*{\parindent}\label{Section_Explicit_Gamma}

Let us calculate the considered contribution to the anomalous dimension explicitly for the simplest regulator function

\begin{equation}\label{F_Simplest}
F(k^2/\Lambda^2) = 1+ k^2/\Lambda^2.
\end{equation}

\noindent
According to Eq. (\ref{Three-Loop_NSVZ}), then we will also obtain the explicit expression for the (considered terms of the) $\beta$-function defined in terms of the bare couplings. Moreover, this calculation allows demonstrating that in the previous section we really deal with the well-defined expressions.

First, we should express the bare Yukawa couplings in terms of the renormalized ones. Due to the absence of divergent quantum corrections to the superpotential \cite{Grisaru:1979wc} the renormalization of the Yukawa couplings is related to the renormalization of the matter superfields. Consequently, it is natural to choose the substraction scheme in which

\begin{equation}\label{Yukawa_Renormalization_General}
\lambda_0^{ijk} = \lambda^{mnp} (Z_\phi^{-1/2})_m{}^i (Z_\phi^{-1/2})_n{}^j (Z_\phi^{-1/2})_p{}^k.
\end{equation}

\noindent
In this paper we calculate a part of the anomalous dimension which does not contain the gauge coupling constant. That is why we are interested only in terms independent of $\alpha$. In the one-loop approximation such terms in the renormalization constant of the matter superfields have the form

\begin{equation}\label{One-Loop_Z}
(Z_\phi)_j{}^i = \delta_j^i - \frac{1}{4\pi^2} \lambda^{imn} \lambda^*_{jmn}\Big(\ln\frac{\Lambda}{\mu} +g_1\Big) + O(\alpha) + O(\lambda^4).
\end{equation}

\noindent
The finite constant $g_1$ appears due to arbitrariness of choosing the subtraction scheme in the considered approximation. Substituting Eq. (\ref{One-Loop_Z}) into Eq. (\ref{Yukawa_Renormalization_General}) we relate the bare Yukawa couplings to the renormalized ones,

\begin{eqnarray}\label{Yukawa_Renormalization}
&& \lambda_0^{ijk} = \lambda^{ijk} + \frac{1}{8\pi^2} \Big(\lambda^{ijm} \lambda^*_{mab} \lambda^{kab} + \lambda^{imk} \lambda^*_{mab} \lambda^{jab} + \lambda^{mjk} \lambda^*_{mab} \lambda^{iab} \Big) \Big(\ln\frac{\Lambda}{\mu} + g_1\Big) \qquad\nonumber\\
&& + O(\alpha\lambda) + O(\lambda^5).
\end{eqnarray}

\noindent
By the help of this equation we express the anomalous dimension (\ref{Gamma_Integral}) (see also (\ref{Complete_Gamma})) in terms of the renormalized Yukawa couplings, on which the derivative with respect to $\ln\Lambda$ does not act,

\begin{eqnarray}\label{Gamma_Via_Renormalized_Lambda}
&&\hspace*{-5mm} \gamma_\phi(\alpha_0,\lambda_0)_j{}^i = \frac{d}{d\ln\Lambda}\Bigg(
\lambda^{iab} \lambda^*_{jab} \int \frac{d^4k}{(2\pi)^4} \frac{2}{k^4 F_k^2}
- 2\lambda^{iab} \lambda^*_{kab}  \lambda^{kcd} \lambda^*_{jcd} \int \frac{d^4k}{(2\pi)^4} \frac{1}{k^4 F_k^2} \nonumber\\
&&\hspace*{-5mm} \times \Big\{\int \frac{d^4l}{(2\pi)^4} \frac{1}{l^4 F_{l}^2} -\frac{1}{4\pi^2} \Big(\ln\frac{\Lambda}{\mu} + g_1\Big) \Big\} - 8\lambda^{iab} \lambda^*_{jac} \lambda^{cde} \lambda^*_{bde} \int \frac{d^4k}{(2\pi)^4} \frac{1}{k^4 F_k^2} \Big\{\int \frac{d^4l}{(2\pi)^4}\nonumber\\
&&\hspace*{-5mm} \times \frac{1}{F_k\, l^2 F_l\, (k+l)^2 F_{k+l}} -\frac{1}{8\pi^2} \Big(\ln\frac{\Lambda}{\mu} + g_1\Big) \Big\}
\Bigg) + O(\alpha) + O(\lambda^6).\vphantom{\Bigg(}\qquad
\end{eqnarray}

\noindent
The term in this expression proportional to $\lambda^{iab} \lambda^*_{kab}  \lambda^{kcd} \lambda^*_{jcd}$ can be easily calculated for an arbitrary function $F(k^2/\Lambda^2)$, such that $F(0)=1$ and $F^{-1}(\infty)=0$. For this purpose we note that the corresponding integral can be presented in the form

\begin{eqnarray}\label{First_Structure}
&& \int \frac{d^4k}{(2\pi)^4} \frac{1}{k^4 F_k^2}  \Big\{\int \frac{d^4l}{(2\pi)^4} \frac{1}{l^4 F_{l}^2} -\frac{1}{4\pi^2} \Big(\ln\frac{\Lambda}{\mu} + g_1\Big) \Big\} = - \frac{1}{64\pi^4} \Big(\ln\frac{\Lambda}{\mu} + g_1\Big)^2\qquad\nonumber\\
&& + \Big[\int \frac{d^4k}{(2\pi)^4} \frac{1}{k^4 F_k^2} -\frac{1}{8\pi^2} \Big(\ln\frac{\Lambda}{\mu} + g_1\Big)\Big]^2.
\end{eqnarray}

\noindent
The second term in Eq. (\ref{First_Structure}) is independent of $\Lambda$. To see this, we take into account that the function $F_k$ depends on $k^2/\Lambda^2$, so that the derivative with respect to $\ln\Lambda$ can be converted into the derivative with respect to $\ln k$ (with the opposite sign). Therefore,

\begin{equation}
\frac{d}{d\ln\Lambda} \Big[\int \frac{d^4k}{(2\pi)^4} \frac{1}{k^4 F_k^2} -\frac{1}{8\pi^2} \Big(\ln\frac{\Lambda}{\mu} + g_1\Big)\Big]
= - \frac{1}{8\pi^2} \int\limits_0^\infty \frac{dk}{k}\,\frac{d}{d\ln k} \Big(\frac{1}{F_k^2}\Big) - \frac{1}{8\pi^2} = 0.
\end{equation}

\noindent
Consequently, the expression (\ref{Gamma_Via_Renormalized_Lambda}) for the considered part of the anomalous dimension can be rewritten as

\begin{eqnarray}\label{Gamma_Final_Integral}
&&\hspace*{-8mm} \gamma_\phi(\alpha_0,\lambda_0)_j{}^i =
\frac{1}{4\pi^2} \lambda^{iab} \lambda^*_{jab}
+ \frac{1}{16\pi^4} \lambda^{iab} \lambda^*_{kab}  \lambda^{kcd} \lambda^*_{jcd} \Big(\ln\frac{\Lambda}{\mu} + g_1\Big) - 8\lambda^{iab} \lambda^*_{jac} \lambda^{cde} \lambda^*_{bde} \frac{d}{d\ln\Lambda}\nonumber\\
&&\hspace*{-8mm} \times \int \frac{d^4k}{(2\pi)^4} \frac{1}{k^4 F_k^2} \Big\{\int \frac{d^4l}{(2\pi)^4} \frac{1}{F_k\, l^2 F_l\, (k+l)^2 F_{k+l}}
 -\frac{1}{8\pi^2} \Big(\ln\frac{\Lambda}{\mu} + g_1\Big) \Big\}
+ O(\alpha) + O(\lambda^6),\qquad
\end{eqnarray}

\noindent
where we take into account that the derivative with respect to $\ln\Lambda$ does not act on the renormalized Yukawa couplings.

For the function $F(k^2/\Lambda^2) = 1+k^2/\Lambda^2$ the remaining integral is calculated in Appendix \ref{Appendix_Explicit_Gamma}. The result obtained there has the form

\begin{eqnarray}
&& \frac{d}{d\ln\Lambda} \int \frac{d^4k}{(2\pi)^4} \frac{1}{k^4 F_k^2} \Big\{\int \frac{d^4l}{(2\pi)^4} \frac{1}{F_k\, l^2 F_l\, (k+l)^2 F_{k+l}} -\frac{1}{8\pi^2} \Big(\ln\frac{\Lambda}{\mu} + g_1\Big) \Big\}\qquad\nonumber\\
&& = \frac{1}{64\pi^4}\Big[\,\frac{1}{2} - \Big(\ln\frac{\Lambda}{\mu}+g_1\Big)\Big].
\end{eqnarray}

\noindent
This implies that the anomalous dimension defined in terms of the bare couplings is given by the expression

\begin{eqnarray}
&&\hspace*{-5mm} \gamma_\phi(\alpha_0,\lambda_0)_j{}^i =
\frac{1}{4\pi^2} \lambda^{iab} \lambda^*_{jab}
+ \frac{1}{16\pi^4} \lambda^{iab} \lambda^*_{kab}  \lambda^{kcd} \lambda^*_{jcd} \Big(\ln\frac{\Lambda}{\mu} + g_1\Big) \nonumber\\
&&\hspace*{-5mm} + \frac{1}{16\pi^4}\lambda^{iab} \lambda^*_{jac} \lambda^{cde} \lambda^*_{bde} \Big[-1 + 2\Big(\ln\frac{\Lambda}{\mu}+g_1\Big)\Big]
+ O(\alpha) + O(\lambda^6).\qquad
\end{eqnarray}

\noindent
The right hand side of this equation depends on the renormalized Yukawa couplings $\lambda$ and $\ln\Lambda/\mu$. Certainly, it should be expressed in terms of the bare Yukawa couplings $\lambda_0$ by the help of Eq. (\ref{Yukawa_Renormalization}). This gives the final result for the considered part of the anomalous dimension,

\begin{equation}\label{Gamma_Result}
\gamma_\phi(\alpha_0,\lambda_0)_j{}^i = \frac{1}{4\pi^2} \lambda_0^{iab} \lambda^*_{0jab}
- \frac{1}{16\pi^4}\lambda_0^{iab} \lambda^*_{0jac} \lambda_0^{cde} \lambda^*_{0bde} + O(\alpha_0) + O(\lambda_0^6).
\end{equation}

\noindent
We see that all $\ln\Lambda/\mu$ disappear. This can be considered as a check of the calculation correctness. Moreover, the finite constant $g_1$, which (partially) determines the subtraction scheme in the one-loop approximation, does not enter the expression for $\gamma_\phi(\alpha_0,\lambda_0)_j{}^i$. This follows from the statement that the RG functions defined in terms of the bare coupling constant are scheme independent for a fixed regularization \cite{Kataev:2013eta}.

The result for the $\beta$-function defined in terms of the bare couplings can be easily found by the help of Eqs. (\ref{Total_Beta}), (\ref{Three-Loop_NSVZ}), and (\ref{Gamma_Result}). Namely, for the regulator (\ref{F_Simplest}) in the considered approximation we obtain

\begin{eqnarray}
&& \frac{\beta(\alpha_0,\lambda_0)}{\alpha_0^2} = -\frac{1}{2\pi}\Big(3C_2-T(R)\Big) -\frac{1}{2\pi r} C(R)_i{}^j \Big(\frac{1}{4\pi^2} \lambda_0^{iab} \lambda^*_{0jab}
- \frac{1}{16\pi^4}\lambda_0^{iab} \lambda^*_{0jac} \lambda_0^{cde} \lambda^*_{0bde}\Big)\qquad\nonumber\\
&& + O(\alpha_0) + O(\lambda_0^6).
\end{eqnarray}

Finally, it should be mentioned that the explicit result obtained for the considered part of the anomalous dimension demonstrates that we really deal with the well-defined expressions.

\section{The NSVZ scheme}
\hspace*{\parindent}\label{Section_NSVZ}

In this section we construct the RG functions defined in terms of the renormalized couplings assuming that the regulator is chosen in the form (\ref{F_Simplest}). The terms of the considered structure in the NSVZ relation are scheme-dependent, so that the NSVZ relation is satisfied only in special subtraction schemes which presumably include the one given by the boundary conditions (\ref{Scheme_Prescription}). Therefore, the purpose of this section is to verify this statement by an explicit calculation.

As a starting point, we integrate the RG equation (\ref{RG_Gamma_Equation}). The result has the form

\begin{eqnarray}
&&\hspace*{-5mm} (\ln Z_\phi)_j{}^i = - \frac{1}{4\pi^2} \lambda^{iab} \lambda^*_{jab} \Big(\ln\frac{\Lambda}{\mu} + g_1\Big)
- \frac{1}{32\pi^4} \lambda^{iab} \lambda^*_{kab}  \lambda^{kcd} \lambda^*_{jcd} \Big(\ln^2\frac{\Lambda}{\mu} + 2 g_1 \ln\frac{\Lambda}{\mu} + \widetilde g_2\Big) \nonumber\\
&&\hspace*{-5mm} - \frac{1}{16\pi^4}\lambda^{iab} \lambda^*_{jac} \lambda^{cde} \lambda^*_{bde} \Big(-\ln\frac{\Lambda}{\mu} + \ln^2\frac{\Lambda}{\mu}+2g_1\ln\frac{\Lambda}{\mu} + g_2 \Big) + O(\alpha) + O(\lambda^6),
\end{eqnarray}

\noindent
where $g_1$, $g_2$, and $\widetilde g_2$ are finite constants. Fixing these constants one fixes the subtraction scheme. To obtain the considered part of the anomalous dimension defined in terms of the renormalized Yukawa couplings, first, it is necessary to express $\ln Z_\phi$ in terms of the bare Yukawa couplings $\lambda_0$ by the help of Eq. (\ref{Yukawa_Renormalization}),

\begin{eqnarray}\label{Z_Constant}
&&\hspace*{-9mm} (\ln Z_\phi)_j{}^i = - \frac{1}{4\pi^2} \lambda_0^{iab} \lambda^*_{0jab} \Big(\ln\frac{\Lambda}{\mu} + g_1\Big)
+ \frac{1}{32\pi^4} \lambda_0^{iab} \lambda^*_{0kab}  \lambda_0^{kcd} \lambda^*_{0jcd} \Big(\ln^2\frac{\Lambda}{\mu} + 2 g_1 \ln\frac{\Lambda}{\mu} +2g_1^2  \nonumber\\
&&\hspace*{-9mm} - \widetilde g_2\Big) + \frac{1}{16\pi^4}\lambda_0^{iab} \lambda^*_{0jac} \lambda_0^{cde} \lambda^*_{0bde} \Big(\ln\frac{\Lambda}{\mu} + \ln^2\frac{\Lambda}{\mu}+2g_1\ln\frac{\Lambda}{\mu} + 2g_1^2 - g_2 \Big) + O(\alpha_0) + O(\lambda_0^6).
\end{eqnarray}

\noindent
Then the contribution to the anomalous dimension

\begin{equation}
\widetilde\gamma_\phi(\alpha,\lambda)_j{}^i = \frac{d(\ln Z_\phi)_j{}^i}{d\ln\mu}
\end{equation}

\noindent
is calculated by differentiating Eq. (\ref{Z_Constant}) with respect to $\ln\mu$ at fixed values of the bare Yukawa couplings $\lambda_0$. This gives

\begin{eqnarray}
&&\hspace*{-5mm} \widetilde\gamma_\phi(\alpha,\lambda)_j{}^i = \frac{1}{4\pi^2} \lambda_0^{iab} \lambda^*_{0jab}
- \frac{1}{16\pi^4} \lambda_0^{iab} \lambda^*_{0kab}  \lambda_0^{kcd} \lambda^*_{0jcd} \Big(\ln\frac{\Lambda}{\mu} + g_1\Big) \nonumber\\
&&\hspace*{-5mm}  + \frac{1}{16\pi^4}\lambda_0^{iab} \lambda^*_{0jac} \lambda_0^{cde} \lambda^*_{0bde} \Big(-1 -2 \ln\frac{\Lambda}{\mu}-2g_1 \Big) + O(\alpha_0) + O(\lambda_0^6).\qquad
\end{eqnarray}

\noindent
The right hand side of this equation should be expressed in terms of the renormalized Yukawa couplings again using Eq. (\ref{Yukawa_Renormalization}),

\begin{equation}\label{Gamma_Renormalized_Result}
\widetilde\gamma_\phi(\alpha,\lambda)_j{}^i = \frac{1}{4\pi^2} \lambda^{iab} \lambda^*_{jab}
- \frac{1}{16\pi^4}\lambda^{iab} \lambda^*_{jac} \lambda^{cde} \lambda^*_{bde} + O(\alpha) + O(\lambda^6).\nonumber\\
\end{equation}

\noindent
We see that this expression does not explicitly depend on $\ln\Lambda/\mu$ that confirms correctness of the calculation. Let us also note that the expression (\ref{Gamma_Renormalized_Result}) is independent of the finite constant $g_1$ which determines the subtraction scheme in the lowest approximation. This implies that the terms of the considered structure in the anomalous dimension are scheme independent. Consequently, Eq. (\ref{Gamma_Renormalized_Result}) should coincide with the corresponding result obtained in the $\overline{\mbox{DR}}$ scheme (see \cite{Jack:1996vg} and references therein). Our notations $\lambda^{ijk}$, $\alpha$, $\widetilde\gamma_\phi(\alpha,\lambda)$, $\widetilde\beta(\alpha,\lambda)$ are related to the corresponding notations of Ref. \cite{Jack:1996vg} $Y^{ijk}$, $g$, $\gamma(g,Y)$, and $\beta_g(g,Y)$ as follows:

\begin{equation}\label{Jack_Jones_Notation}
\lambda^{ijk} = \frac{1}{2} Y^{ijk};\qquad \alpha = \frac{g^2}{4\pi};\qquad \widetilde\gamma_\phi(\alpha,\lambda)=2\gamma(g,Y);\qquad \widetilde\beta(\alpha,\lambda) = \frac{g\beta_g(g,Y)}{2\pi}.
\end{equation}

\noindent
Using these equations one can easily verify that the terms of the considered structure in Ref. \cite{Jack:1996vg} agree with Eq. (\ref{Gamma_Renormalized_Result}).

Now, let us proceed to calculating the $\beta$-function defined in terms of the renormalized couplings. We start with integrating the RG equation

\begin{equation}
\frac{d}{d\ln\Lambda}\Big(\frac{1}{\alpha_0}\Big) = - \frac{\beta(\alpha_0,\lambda_0)}{\alpha_0^2},
\end{equation}

\noindent
taking into consideration the one-loop result (see Ref. \cite{Aleshin:2016yvj}), the two-loop terms quadratic in the Yukawa couplings, and the three-loop terms quartic in the Yukawa couplings. Then we obtain the equation relating the bare coupling constant to the renormalized one,

\begin{eqnarray}\label{Bare_Via_Renormalized}
&& \frac{1}{\alpha_0} = \frac{1}{\alpha} +\frac{1}{2\pi}\Big(3C_2-T(R)\Big)\Big(\ln\frac{\Lambda}{\mu} + b_1\Big) +\frac{1}{2\pi r} C(R)_i{}^j \Big[
\, \frac{1}{4\pi^2} \lambda^{iab} \lambda^*_{jab} \Big(\ln\frac{\Lambda}{\mu} + b_2\Big)\qquad\nonumber\\
&& + \frac{1}{32\pi^4} \lambda^{iab} \lambda^*_{kab}  \lambda^{kcd} \lambda^*_{jcd} \Big(\ln^2\frac{\Lambda}{\mu}
+ 2 g_1 \ln\frac{\Lambda}{\mu} + \widetilde b_3\Big) + \frac{1}{16\pi^4}\lambda^{iab} \lambda^*_{jac} \lambda^{cde} \lambda^*_{bde} \Big(-\ln\frac{\Lambda}{\mu}\nonumber\\
&& + \ln^2\frac{\Lambda}{\mu}+2g_1\ln\frac{\Lambda}{\mu} + b_3 \Big)\Big] + O(\alpha) + O(\lambda^6),
\end{eqnarray}

\noindent
where $b_1$, $b_2$, $b_3$, and $\widetilde b_3$ are arbitrary finite constants determining the subtraction scheme in the considered approximation. Certainly, in the three-loop approximation there are also terms proportional to $\alpha$ (a part of the two-loop contribution), $\alpha^2$, and $\alpha\lambda^2$. However, in this paper we do not consider them.

At the next step, we solve Eq. (\ref{Bare_Via_Renormalized}) for the renormalized coupling constant $\alpha$ and write the result in terms of the bare gauge and Yukawa couplings by the help of Eq. (\ref{Yukawa_Renormalization}). In the considered approximation the result is written as

\begin{eqnarray}\label{Renormalized_Via_Bare}
&& \frac{1}{\alpha} = \frac{1}{\alpha_0} - \frac{1}{2\pi}\Big(3C_2-T(R)\Big)\Big(\ln\frac{\Lambda}{\mu} + b_1\Big) -\frac{1}{2\pi r} C(R)_i{}^j \Big[\,
\frac{1}{4\pi^2} \lambda_0^{iab} \lambda^*_{0jab} \Big(\ln\frac{\Lambda}{\mu} + b_2\Big)\qquad
\nonumber\\
&& - \frac{1}{32\pi^4} \lambda_0^{iab} \lambda^*_{0kab}  \lambda_0^{kcd} \lambda^*_{0jcd} \Big(\ln^2\frac{\Lambda}{\mu}
+ 2 b_2 \ln\frac{\Lambda}{\mu} + 2 b_2 g_1 - \widetilde b_3\Big) - \frac{1}{16\pi^4}\lambda_0^{iab} \lambda^*_{0jac} \lambda_0^{cde} \lambda^*_{0bde}
\nonumber\\
&& \times \Big(\ln\frac{\Lambda}{\mu} + \ln^2\frac{\Lambda}{\mu}+2b_2\ln\frac{\Lambda}{\mu} + 2 b_2 g_1 - b_3 \Big)\Big] + O(\alpha_0) + O(\lambda_0^6).
\end{eqnarray}

\noindent
Differentiating $1/\alpha$ with respect to $\ln\mu$ at fixed values of the bare gauge and Yukawa couplings, we obtain the $\beta$-function defined in terms of the renormalized constants,

\begin{equation}
\frac{\widetilde\beta(\alpha,\lambda)}{\alpha^2} = - \frac{d}{d\ln\mu}\Big(\frac{1}{\alpha}\Big).
\end{equation}

\noindent
A part of this $\beta$-function corresponding to the terms of the considered structure, which is obtained from the derivative of Eq. (\ref{Renormalized_Via_Bare}), has the form

\begin{eqnarray}
&& \frac{\widetilde\beta(\alpha,\lambda)}{\alpha^2} = -\frac{1}{2\pi}\Big(3C_2 - T(R)\Big) + \frac{1}{2\pi r} C(R)_i{}^j \Big[ -\frac{1}{4\pi^2} \lambda_0^{iab} \lambda^*_{0jab} + \frac{1}{16\pi^4} \lambda_0^{iab} \lambda^*_{0kab}  \lambda_0^{kcd} \lambda^*_{0jcd}\qquad\nonumber\\
&& \times \Big(\ln\frac{\Lambda}{\mu} + b_2 \Big) + \frac{1}{16\pi^4}\lambda_0^{iab} \lambda^*_{0jac} \lambda_0^{cde} \lambda^*_{0bde} \Big(1 + 2 \ln\frac{\Lambda}{\mu}+2b_2 \Big)\Big] + O(\alpha_0) + O(\lambda_0^6).
\end{eqnarray}

\noindent
As usual, the right hand side should be expressed in terms of the renormalized Yukawa couplings by the help of Eq. (\ref{Yukawa_Renormalization}). This gives the final result for the considered part of the $\beta$-function,

\begin{eqnarray}\label{Final_Beta}
&& \frac{\widetilde\beta(\alpha,\lambda)}{\alpha^2} =  -\frac{1}{2\pi}\Big(3C_2 - T(R)\Big) + \frac{1}{2\pi r} C(R)_i{}^j \Big[ -\frac{1}{4\pi^2} \lambda^{iab} \lambda^*_{jab} + \frac{1}{16\pi^4} \lambda^{iab} \lambda^*_{kab}  \lambda^{kcd} \lambda^*_{jcd}\qquad\nonumber\\
&& \times \Big(b_2 -g_1\Big)
+ \frac{1}{16\pi^4}\lambda^{iab} \lambda^*_{jac} \lambda^{cde} \lambda^*_{bde} \Big(1 + 2b_2 -2g_1\Big)\Big] + O(\alpha) + O(\lambda^6).\qquad
\end{eqnarray}

\noindent
We see that this expression contains the constants $b_2$ and $g_1$ and is, therefore, scheme-dependent. Note that it is written in an arbitrary scheme, so that the result obtained in $\overline{\mbox{DR}}$-scheme should be a particular case of Eq. (\ref{Final_Beta}). (The results obtained with various regularizations can be related by a specially tuned finite renormalization or, equivalently, by a special choice of the finite constants defining the subtraction scheme.) The $\overline{\mbox{DR}}$ result has been obtained in \cite{Jack:1996vg}. It can be written in the notation of this paper via Eq. (\ref{Jack_Jones_Notation}) as

\begin{eqnarray}\label{DR_Beta}
&&\hspace*{-5mm} \widetilde\beta_{\overline{\mbox{\scriptsize DR}}}(\alpha,\lambda) = -\frac{\alpha^2}{2\pi}\Big(3C_2 - T(R)\Big) + \frac{\alpha^2}{2\pi r} C(R)_i{}^j  \Big[ - \frac{1}{4\pi^2} \lambda^*_{jab} \lambda^{iab} + \frac{1}{64 \pi^4}  \Big(\lambda^{iab} \lambda^*_{kab} \lambda^{kcd} \lambda^*_{jcd}\nonumber\\
&&\hspace*{-5mm} + 6 \lambda^{iab} \lambda^*_{jac} \lambda^{cde} \lambda^*_{bde} \Big)\Big] + O(\alpha^3) + O(\alpha^2\lambda^6).\vphantom{\frac{1}{2}}
\end{eqnarray}

\noindent
Comparing Eqs. (\ref{Final_Beta}) and (\ref{DR_Beta}), we see that they coincide for

\begin{equation}\label{DR_Constants}
b_2-g_1 =\frac{1}{4}.
\end{equation}

\noindent
This implies that our results agree with the results of \cite{Jack:1996vg}, certainly, taking into account that the regularizations and the subtraction schemes are different. Also it is easy to see \cite{Jack:1996vg} that for the finite constants satisfying Eq. (\ref{DR_Constants}) the NSVZ relation is not valid.

Next, let us verify that the prescription (\ref{Scheme_Prescription}), proposed in \cite{Stepanyantz:2016gtk}, really gives the NSVZ scheme. First, we compare Eqs. (\ref{Gamma_Renormalized_Result}) and (\ref{Final_Beta}) and note that the NSVZ relation is not valid in an arbitrary subtraction scheme (which is defined by the coefficients $b$ and $g$).

Then, let us impose the boundary condition $Z_\phi(\alpha,\lambda,x_0)_i{}^j = \delta_i{}^j$. Substituting $\ln\Lambda/\mu$ by the fixed value $x_0$ in the expression $(Z_\phi)_i{}^j$ we solve the above equation for the finite constants $g_1$ etc. In the lowest approximation this gives $g_1 = - x_0$. Similarly, we find the constants $b_1$, $b_2$ etc. from the boundary condition $Z_\alpha(\alpha,\lambda,x_0) = \alpha/\alpha_0 =1$. Namely, we solve the equation $1/\alpha = 1/\alpha_0$ with $\ln\Lambda/\mu=x_0$ for the constants $b$. The result has the form $b_1 = b_2 = -x_0$. This implies that in the scheme defined by the prescription (\ref{Scheme_Prescription})

\begin{equation}
b_2 - g_1 = 0.
\end{equation}

\noindent
Consequently, in the scheme (\ref{Scheme_Prescription})

\begin{eqnarray}
&& \frac{\widetilde\beta(\alpha,\lambda)}{\alpha^2} =  -\frac{1}{2\pi}\Big(3C_2 - T(R)\Big) + \frac{1}{2\pi r} C(R)_i{}^j \Big[ -\frac{1}{4\pi^2} \lambda^{iab} \lambda^*_{jab} + \frac{1}{16\pi^4}\lambda^{iab} \lambda^*_{jac} \lambda^{cde} \lambda^*_{bde} \Big]\qquad\nonumber\\
&& + O(\alpha) + O(\lambda^6) = -\frac{1}{2\pi}\Big(3C_2 - T(R)\Big) - \frac{1}{2\pi r} C(R)_i{}^j \widetilde\gamma_\phi(\alpha,\lambda)_i{}^j + O(\alpha) + O(\lambda^6).\qquad
\end{eqnarray}

\noindent
Thus, under the condition (\ref{Scheme_Prescription}) the NSVZ relation is satisfied for terms of the considered structure. This confirms the guess made in \cite{Stepanyantz:2016gtk}.

\section{Conclusion}
\hspace*{\parindent}

In this paper we have verified the relation between the two-point Green functions of ${\cal N}=1$ SYM for the contributions quartic in the Yukawa couplings in the case of using the higher covariant derivative regularization. For this regularization it was demonstrated that (in the considered approximation and for the terms of the considered structure) the NSVZ relation is satisfied by the RG functions defined in terms of the bare couplings as it was suggested in \cite{Stepanyantz:2016gtk}. Exactly as in the Abelian case, this follows from the factorization of the loop integrals into integrals of double total derivatives in the momentum space. Consequently, it is possible to calculate one of these integrals and relate the three-loop contribution to the $\beta$-function to the two-loop contribution to the anomalous dimension. For the RG functions defined in terms of the renormalized couplings, we have checked that the prescription proposed in \cite{Stepanyantz:2016gtk} really gives the NSVZ scheme. It should be noted that this check is not trivial, because the considered terms in the NSVZ relation are scheme dependent. Thus, we confirmed the proposals made in \cite{Stepanyantz:2016gtk} by the explicit calculations.

\appendix

\section{Explicit expressions for the diagrams}
\hspace{\parindent}\label{Appendix_Explicit_Graphs}

In this section we present the results for all supergraphs shown in Figs. \ref{Figure_Beta_Two_Loop} and \ref{Figure_Beta}. In the Minkowski space the result for any supergraph contributing to the two-point Green function of the background gauge superfield can be written in the form

\begin{equation}
\Delta\Gamma = \int \frac{d^4p}{(2\pi)^4}\, d^4\theta\,\Big[ \bm{V}(p,\theta)_i{}^j \partial^2\Pi_{1/2} \bm{V}(-p,\theta)_k{}^l I_{\mbox{\scriptsize inv}}(p)_{jl}{}^{ik} + \bm{V}(p,\theta)_i{}^j \bm{V}(-p,\theta)_k{}^l I_{\mbox{\scriptsize non-inv}}(p)_{jl}{}^{ik} \Big],
\end{equation}

\noindent
where $\Delta\Gamma$ is the corresponding contribution to the effective action. Due to the background gauge invariance the non-invariant terms cancel each other in the sum of all superdiagrams,

\begin{equation}\label{Non-Invariant_Cancellation}
\sum\limits_{\mbox{\scriptsize all supergraphs}} (T^A)_i{}^j (T^B)_k{}^l\, I_{\mbox{\scriptsize non-inv}}(p)_{jl}{}^{ik} = 0.
\end{equation}

\noindent
The sum of the invariant terms determines the function $d^{-1}-\alpha_0^{-1}$ according to Eq. (\ref{Two_Point_Function}). To write the result in the most convenient form, we note that $(T^A)_i{}^j (T^B)_k{}^l \big(I_{\mbox{\scriptsize inv}}\big)_{jl}{}^{ik}$ is the invariant tensor. In this paper we consider simple gauge groups, for which it should be proportional to $\delta^{AB}$. Therefore,

\begin{equation}\label{Generators_Result}
(T^A)_i{}^j (T^B)_k{}^l \big(I_{\mbox{\scriptsize inv}}\big)_{jl}{}^{ik} = \frac{1}{r} \delta^{AB}\, (T^C)_i{}^j (T^C)_k{}^l \big(I_{\mbox{\scriptsize inv}}\big)_{jl}{}^{ik}.
\end{equation}

\noindent
Thus, from Eq. (\ref{Two_Point_Function}) we obtain

\begin{equation}\label{D-1}
d^{-1}(\alpha_0,\lambda_0,\Lambda/p) - \alpha_0^{-1} = -\frac{16\pi}{r} (T^C)_i{}^j (T^C)_k{}^l \sum\limits_{\mbox{\scriptsize all supergraphs}} I_{\mbox{\scriptsize inv}}(p)_{jl}{}^{ik}.
\end{equation}

\noindent
We are interested in the derivative of this function with respect to $\ln\Lambda$ in the limit of the vanishing external momentum. That is why we can calculate the functions $(I_{\mbox{\scriptsize inv}})_{jl}{}^{ik}$ and $(I_{\mbox{\scriptsize non-inv}})_{jl}{}^{ik}$ in the limit $p\to 0$. Certainly, in this case expressions for individual supergraphs are not well-defined. However, the sum of invariant contributions differentiated with respect to $\ln\Lambda$ is well-defined due to Eq. (\ref{D_Derivative}).

Below we present expressions for the functions $(I_{\mbox{\scriptsize inv}})_{jl}{}^{ik}$ and $(I_{\mbox{\scriptsize non-inv}})_{jl}{}^{ik}$ in the limit $p\to 0$ for all supergraphs in Figs. \ref{Figure_Beta_Two_Loop} and \ref{Figure_Beta} in the form

\begin{equation}
\mbox{Supergraph} = \bm{V}_i{}^j \partial^2\Pi_{1/2} \bm{V}_k{}^l I_{\mbox{\scriptsize inv}}(p=0)_{jl}{}^{ik} + \bm{V}_i{}^j \bm{V}_k{}^l I_{\mbox{\scriptsize non-inv}}(p=0)_{jl}{}^{ik},
\end{equation}

\noindent
where the coefficients $I_{jl}{}^{ik}$ are written as integrals over Euclidean momentums which are obtained after the Wick rotation. Using these expressions one can verify Eq. (\ref{Non-Invariant_Cancellation}) in the limit $p\to 0$ and obtain the function (\ref{D-1}), which, after differentiating with respect to $\ln\Lambda$, gives the $\beta$-function defined in terms of the bare couplings. In the equations presented below the prime denotes the derivative with respect to the square of the momentum,

\begin{equation}
F_k' \equiv \frac{d}{dk^2} F(k^2/\Lambda^2).
\end{equation}

Let us start with the supergraphs presented in Fig. \ref{Figure_Beta_Two_Loop}. They are given by the following expressions:

\begin{eqnarray}
&& (1) = \lambda_0^{ika} \lambda^*_{0jla} \int \frac{d^4q}{(2\pi)^4} \frac{d^4k}{(2\pi)^4} \frac{1}{q^4 F_q^2\, k^4 F_k^2\, (q+k)^2 F_{q+k}} \Big[\Big((q+k)^2 F_k F_q + 2k^2 (2q_\mu k^\mu  \nonumber\\
&&\quad\ \ + q^2) F_q F_k' + 2k^2 q^2 q_\mu k^\mu F_k' F_q'\Big) \bm{V}_i{}^j \partial^2\Pi_{1/2}\bm{V}_k{}^l + 2 q^2 F_q\, k^2 F_k\, \bm{V}_i{}^j \bm{V}_k{}^l \Big];\vphantom{\frac{d^4k}{(2\pi)^4}}\\
&& (2) = \lambda_0^{iab} \lambda^*_{0jab} \int \frac{d^4q}{(2\pi)^4} \frac{d^4k}{(2\pi)^4} \frac{1}{q^4 F_q^3\, k^2 F_k\, (q+k)^2 F_{q+k}} \Big[\Big(F_q^2 + 2 q^2 F_q' F_q + 2 q^4 (F_q')^2\Big) \nonumber\\
&&\quad\ \ \times  \big(\bm{V} \partial^2\Pi_{1/2}\bm{V}\big)_i{}^j + 2 q^2 F_q^2 \big(\bm{V}^2\big)_i{}^j\Big];\vphantom{\frac{d^4k}{(2\pi)^4}}\\
&& (3) = - \lambda_0^{iab} \lambda^*_{0jab} \int \frac{d^4q}{(2\pi)^4} \frac{d^4k}{(2\pi)^4} \frac{1}{q^2 F_q^2\, k^2 F_k\, (q+k)^2 F_{q+k}} \Big[(q^2 F_q'' +F_q') \big(\bm{V} \partial^2\Pi_{1/2}\bm{V}\big)_i{}^j \qquad\nonumber\\
&&\quad\ \ + F_q \big(\bm{V}^2\big)_i{}^j\Big].\qquad\vphantom{\frac{d^4k}{(2\pi)^4}}
\end{eqnarray}

\noindent
To find the sum of these diagrams, it is necessary to take into account the identity

\begin{equation}\label{Yukawa_Quadratic_Identity}
\lambda^*_{0jla} \lambda_0^{ika} (T^A)_k{}^l = - \frac{1}{2} \lambda^*_{0jab}\lambda_0^{kab} (T^A)_k{}^i = -\frac{1}{2} (T^A)_j{}^k \lambda^*_{0kab} \lambda_0^{iab},
\end{equation}

\noindent
which follows from Eq. (\ref{Yukawa_Identity}). Rewriting the expression for the diagram (1) by the help of Eq. (\ref{Yukawa_Quadratic_Identity}), we obtain that the non-invariant terms cancel each other, and the sum of the invariant terms is

\begin{equation}
\frac{1}{8} \lambda_0^{iab} \lambda^*_{0jab} \big(\bm{V}\partial^2\Pi_{1/2} \bm{V}\big)_i{}^j \int \frac{d^4q}{(2\pi)^4} \frac{d^4k}{(2\pi)^4} \frac{\partial}{\partial q^\mu} \frac{\partial}{\partial q_\mu} \Big(\frac{1}{k^2 F_k\, q^2 F_q\, (q+k)^2 F_{q+k}}\Big).
\end{equation}

\noindent
Consequently, the contribution to the function $d^{-1}-\alpha_0^{-1}$ from the considered (two-loop) diagrams can be written as the integral of the double total derivative

\begin{equation}\label{Two-Loop_Result}
- \frac{2\pi}{r} C(R)_i{}^j \lambda^*_{0jab} \lambda_0^{iab} \int \frac{d^4q}{(2\pi)^4} \frac{d^4k}{(2\pi)^4} \frac{\partial}{\partial q^\mu} \frac{\partial}{\partial q_\mu} \Big(\frac{1}{k^2 F_k\, q^2 F_q\, (q+k)^2 F_{q+k}}\Big).
\end{equation}

The supergraphs presented in Fig. \ref{Figure_Beta} are given by the following expressions:

\begin{eqnarray}\label{Diagram1}
&&\hspace*{-5mm} (1) = \lambda_0^{iab} \lambda^*_{0jad} \lambda_0^{dek} \lambda^*_{0bel} \int \frac{d^4q}{(2\pi)^4} \frac{d^4k}{(2\pi)^4} \frac{d^4l}{(2\pi)^4} \frac{4}{q^4 F_q^2\, k^4 F_k^2\, l^4 F_l^2\, (q+k)^2 F_{q+k}\,(k+l)^2 F_{k+l}}\nonumber\\
&&\hspace*{-5mm} \quad\ \ \times \Big[ \Big(-2q^2 k^2 l^2 q_\mu l^\mu F_q' F_l' + 2 l^2 F_q F_l' \big( -q^2 (k+l)^2 + q^2 l^2 -2 k^2 q_\mu l^\mu \big) + F_q F_l \big( -2 q^2 (k+l)^2 \vphantom{\frac{d^4k}{(2\pi)^4}}\nonumber\\
&&\hspace*{-5mm} \quad\ \ + q^2 l^2 -2 k^2 q_\mu l^\mu\big)\Big)\bm{V}_i{}^j \partial^2\Pi_{1/2}\bm{V}_k{}^l - 2 q^2 F_q\, k^2 l^2 F_l\, \bm{V}_i{}^j \bm{V}_k{}^l\Big];\vphantom{\frac{d^4k}{(2\pi)^4}}\\
\label{Diagram2}
&&\hspace*{-5mm}  (2) = - \lambda_0^{iab} \lambda^*_{0lab} \lambda_0^{kcd} \lambda^*_{0jcd} \int \frac{d^4q}{(2\pi)^4} \frac{d^4k}{(2\pi)^4} \frac{d^4l}{(2\pi)^4} \frac{1}{q^2 F_q\, k^4 F_k^4\, l^2 F_l\, (q+k)^2 F_{q+k}\, (k+l)^2 F_{k+l}}\nonumber\\
&&\hspace*{-5mm} \quad\ \ \times \Big[\Big(F_k^2 + 2 k^2 F_k' F_k + 2 k^4 (F_k')^2\Big) \bm{V}_i{}^j \partial^2\Pi_{1/2}\bm{V}_k{}^l + 2 k^2 F_k^2 \bm{V}_i{}^j \bm{V}_k{}^l \Big];\vphantom{\frac{d^4k}{(2\pi)^4}}\\
\label{Diagram3}
&& \hspace*{-5mm}  (3) = - \lambda_0^{ika} \lambda^*_{0jlb} \lambda_0^{bcd} \lambda^*_{0acd} \int \frac{d^4q}{(2\pi)^4} \frac{d^4k}{(2\pi)^4} \frac{d^4l}{(2\pi)^4} \frac{2}{q^4 F_q^2\, k^2 F_k^2\, l^2 F_l\, (q+k)^4 F_{q+k}^2\, (k+l)^2 F_{k+l}}\nonumber\\
&&\hspace*{-5mm} \quad\ \ \times \Big[\Big(k^2 F_q F_{q+k} + 2 q^2 (k^2 -q^2) F_q' F_{q+k} + q^2 (q+k)^2 \big(k^2-2q^2\big) F'_q F'_{q+k}\Big) \bm{V}_i{}^j \partial^2\Pi_{1/2} \bm{V}_k{}^l\vphantom{\frac{d^4k}{(2\pi)^4}}\nonumber\\
&&\hspace*{-5mm} \quad\ \  + 2 q^2 F_q\, (q+k)^2 F_{q+k} \bm{V}_i{}^j \bm{V}_k{}^l \Big];\vphantom{\frac{d^4k}{(2\pi)^4}}\\
\label{Diagram4}
&&\hspace*{-5mm}  (4) = \lambda_0^{ikb} \lambda^*_{0alb} \lambda_0^{acd} \lambda^*_{0jcd} \int \frac{d^4q}{(2\pi)^4} \frac{d^4k}{(2\pi)^4} \frac{d^4l}{(2\pi)^4} \frac{8}{q^4 F_q^2\, k^4 F_k^3\, l^2 F_l\, (q+k)^2 F_{q+k}\, (k+l)^2 F_{k+l}}\nonumber\\
&&\hspace*{-5mm} \quad\ \ \times \Big[\Big(-2 q^2 k^2 q_\mu k^\mu F_q' F_k' -k^2 F_k' F_q \big((k+q)^2 - k^2\big) -q^2 F_q' F_k \big((k+q)^2 - q^2\big) - (k+q)^2  \vphantom{\frac{d^4k}{(2\pi)^4}}\nonumber\\
&&\hspace*{-5mm} \quad\ \ \times F_q F_k \Big) \bm{V}_i{}^j \partial^2\Pi_{1/2} \bm{V}_k{}^l - 2 q^2 F_q\, k^2F_k\,\bm{V}_i{}^j \bm{V}_k{}^l\Big];\vphantom{\frac{d^4k}{(2\pi)^4}}\\
\label{Diagram5}
&&\hspace*{-5mm}  (5) = -\lambda_0^{iab} \lambda^*_{0kab} \lambda_0^{kcd} \lambda^*_{0jcd} \int \frac{d^4q}{(2\pi)^4} \frac{d^4k}{(2\pi)^4} \frac{d^4l}{(2\pi)^4} \frac{2}{q^2 F_q\, k^4 F_k^4\, l^2 F_l\, (q+k)^2 F_{q+k}\, (k+l)^2 F_{k+l}}\nonumber\\
&&\hspace*{-5mm} \quad\ \ \times \Big[\Big(F_k^2 + 2 k^2 F_k' F_k + 2 k^4 (F_k')^2\Big) \big(\bm{V} \partial^2\Pi_{1/2}\bm{V}\big)_i{}^j + 2 k^2 F_k^2 \big(\bm{V}^2\big)_i{}^j\Big];\vphantom{\frac{d^4k}{(2\pi)^4}}\\
\label{Diagram6}
&&\hspace*{-5mm}  (6) = - \lambda_0^{iac} \lambda^*_{0jad} \lambda_0^{def} \lambda^*_{0cef} \int \frac{d^4q}{(2\pi)^4} \frac{d^4k}{(2\pi)^4} \frac{d^4l}{(2\pi)^4} \frac{4}{q^4 F_q^3\, k^2 F_k^2\, l^2 F_l\, (q+k)^2 F_{q+k}\, (k+l)^2 F_{k+l}}\nonumber\\
&&\hspace*{-5mm} \quad\ \ \times \Big[\Big(F_q^2 + 2 q^2 F_q' F_q + 2 q^4 (F_q')^2\Big) \big(\bm{V} \partial^2\Pi_{1/2}\bm{V}\big)_i{}^j + 2 q^2 F_q^2 \big(\bm{V}^2\big)_i{}^j\Big];\vphantom{\frac{d^4k}{(2\pi)^4}}\\
\label{Diagram7}
&&\hspace*{-5mm}  (7) = \lambda_0^{iab} \lambda^*_{0kab} \lambda_0^{kcd} \lambda^*_{0jcd} \int \frac{d^4q}{(2\pi)^4} \frac{d^4k}{(2\pi)^4} \frac{d^4l}{(2\pi)^4} \frac{2}{q^2 F_q\, k^2 F_k^3\, l^2 F_l\, (q+k)^2 F_{q+k}\, (k+l)^2 F_{k+l}}\nonumber\\
&&\hspace*{-5mm} \quad\ \ \times \Big[(k^2 F_k'' +F_k') \big(\bm{V} \partial^2\Pi_{1/2}\bm{V}\big)_i{}^j   + F_k \big(\bm{V}^2\big)_i{}^j\Big];\vphantom{\frac{d^4k}{(2\pi)^4}}\\
\label{Diagram8}
&&\hspace*{-5mm}  (8) = \lambda_0^{iac} \lambda^*_{0jad} \lambda_0^{def} \lambda^*_{0cef} \int \frac{d^4q}{(2\pi)^4} \frac{d^4k}{(2\pi)^4} \frac{d^4l}{(2\pi)^4} \frac{4}{q^2 F_q^2\, k^2 F_k^2\, l^2 F_l\, (q+k)^2 F_{q+k}\, (k+l)^2 F_{k+l}}\nonumber\\
&&\hspace*{-5mm} \quad\ \ \times \Big[(q^2 F_q'' +F_q') \big(\bm{V} \partial^2\Pi_{1/2}\bm{V}\big)_i{}^j   + F_q \big(\bm{V}^2\big)_i{}^j\Big].\vphantom{\frac{d^4k}{(2\pi)^4}}
\end{eqnarray}

\noindent
Various structures formed by the Yukawa constants in these expressions can be reduced to two basic combinations by the help of Eq. (\ref{Yukawa_Identity}). For example, the non-invariant terms are proportional to

\begin{eqnarray}\label{Y4_identity1}
&&\hspace*{-7mm} (1) \to \lambda_0^{iab} \lambda^*_{0jad} \lambda_0^{dek} \lambda^*_{0bel} \bm{V}_i{}^j \bm{V}_k{}^l = \frac{1}{4} \lambda_0^{iab} \lambda^*_{0kab} \lambda_0^{kcd} \lambda^*_{0jcd} (\bm{V}^2)_i{}^j;\\
\label{Y4_identity2}
&&\hspace*{-7mm} (2) \to \lambda_0^{iab} \lambda^*_{0lab} \lambda_0^{kcd} \lambda^*_{0jcd} \bm{V}_i{}^j \bm{V}_k{}^l = \lambda_0^{iab} \lambda^*_{0kab} \lambda_0^{kcd} \lambda^*_{0jcd} (\bm{V}^2)_i{}^j;\vphantom{\frac{1}{2}}\\
\label{Y4_identity3}
&&\hspace*{-7mm} (3) \to \lambda_0^{ika} \lambda^*_{0jlb} \lambda_0^{bcd} \lambda^*_{0acd} \bm{V}_i{}^j \bm{V}_k{}^l = \Big(\frac{1}{2} \lambda_0^{iab} \lambda^*_{0kab} \lambda_0^{kcd} \lambda^*_{0jcd} - \lambda_0^{iac} \lambda^*_{0jad} \lambda_0^{def} \lambda^*_{0cef}\Big) (\bm{V}^2)_i{}^j;\\
\label{Y4_identity4}
&&\hspace*{-7mm} (4) \to \lambda_0^{ikb} \lambda^*_{0alb} \lambda_0^{acd} \lambda^*_{0jcd} \bm{V}_i{}^j \bm{V}_k{}^l = -\frac{1}{2} \lambda_0^{iab} \lambda^*_{0kab} \lambda_0^{kcd} \lambda^*_{0jcd} (\bm{V}^2)_i{}^j,
\end{eqnarray}

\noindent
where we take into account that $\bm{V}_i{}^j = e_0 \bm{V}^A (T^A)_i{}^j$.

Using these identities one can verify that all non-invariant terms in the considered three-loop diagrams cancel each other. This fact can be considered as a test of the calculation correctness, because the non-invariant terms should vanish due to the background gauge invariance of the effective action.

Using identities similar to Eqs. (\ref{Y4_identity1}) --- (\ref{Y4_identity4}) for the invariant terms, after some transformations the sum of the expressions Eqs. (\ref{Diagram1}) --- (\ref{Diagram8}) can be presented as the following integral of double total derivatives:

\begin{eqnarray}
&& -\frac{1}{4} \big(\bm{V}\partial^2\Pi_{1/2}\bm{V}\big)_i{}^j \int \frac{d^4k}{(2\pi)^4} \frac{d^4l}{(2\pi)^4} \frac{d^4q}{(2\pi)^4} \Big[ \lambda_0^{iab}\lambda^*_{0kab} \lambda_0^{kcd}\lambda^*_{0jcd} \Big(\frac{\partial}{\partial k_\mu} \frac{\partial}{\partial k^\mu} - \frac{\partial}{\partial q_\mu} \frac{\partial}{\partial q^\mu}\Big)\qquad\nonumber\\
&& + 2 \lambda_0^{iab}\lambda^*_{0jac} \lambda_0^{cde}\lambda^*_{0bde}\, \frac{\partial}{\partial q_\mu} \frac{\partial}{\partial q^\mu} \Big] \frac{1}{k^2 F_k^2\, q^2 F_q\, (q+k)^2 F_{q+k}\, l^2 F_l\, (l+k)^2 F_{l+k}}.
\end{eqnarray}

\noindent
From this expression we obtain that the contribution of the diagrams shown in Fig. \ref{Figure_Beta} to the function $d^{-1}-\alpha_0^{-1}$ is

\begin{eqnarray}\label{Three-loop_Result}
&& \frac{4\pi}{r} C(R)_i{}^j \int \frac{d^4k}{(2\pi)^4} \frac{d^4l}{(2\pi)^4} \frac{d^4q}{(2\pi)^4} \Big[ \lambda_0^{iab}\lambda^*_{0kab} \lambda_0^{kcd}\lambda^*_{0jcd} \Big(\frac{\partial}{\partial k_\mu} \frac{\partial}{\partial k^\mu} - \frac{\partial}{\partial q_\mu} \frac{\partial}{\partial q^\mu}\Big)\qquad\nonumber\\
&& + 2 \lambda_0^{iab}\lambda^*_{0jac} \lambda_0^{cde}\lambda^*_{0bde}\, \frac{\partial}{\partial q_\mu} \frac{\partial}{\partial q^\mu} \Big] \frac{1}{k^2 F_k^2\, q^2 F_q\, (q+k)^2 F_{q+k}\, l^2 F_l\, (l+k)^2 F_{l+k}}.\qquad
\end{eqnarray}

\noindent
Summing Eqs. (\ref{Two-Loop_Result}) and (\ref{Three-loop_Result}) and differentiating the result with respect to $\ln\Lambda$ we obtain Eq. (\ref{Delta_Beta_Integral}).

\section{Calculation of integrals with higher derivatives regularization}
\hspace*{\parindent}\label{Appendix_Explicit_Gamma}

In this appendix we calculate the expression

\begin{equation}\label{Integral_We_Calculate}
I \equiv \frac{d}{d\ln\Lambda} \int \frac{d^4k}{(2\pi)^4} \frac{1}{k^4 F_k^2} \Big\{\int \frac{d^4l}{(2\pi)^4} \frac{1}{F_k\, l^2 F_l\, (k+l)^2 F_{k+l}} -\frac{1}{8\pi^2} \Big(\ln\frac{\Lambda}{\mu} + g_1\Big) \Big\}
\end{equation}

\noindent
entering Eq. (\ref{Gamma_Final_Integral}) for the regulator $F(k^2/\Lambda^2)= 1+k^2/\Lambda^2$. Then the integral over $d^4l$ can be written as

\begin{equation}
\quad\int \frac{d^4l}{(2\pi)^4} \frac{1}{l^2 F_l (k+l)^2 F_{k+l}} = \int \frac{d^4l}{(2\pi)^4} \Big(\frac{1}{l^2} - \frac{1}{l^2 +\Lambda^2}\Big)\Big(\frac{1}{(k+l)^2} - \frac{1}{(k+l)^2+\Lambda^2}\Big) = 2I_1-I_2,\quad
\end{equation}

\noindent
where we introduce the notation

\begin{eqnarray}\label{I1_Definition}
&& I_1 \equiv \int \frac{d^4l}{(2\pi)^4} \Big(\frac{1}{l^2} - \frac{1}{l^2 +\Lambda^2}\Big) \frac{1}{(k+l)^2};\\
\label{I2_Definition}
&& I_2 \equiv \int \frac{d^4l}{(2\pi)^4} \Big(\frac{1}{l^2 (k+l)^2} - \frac{1}{(l^2 +\Lambda^2)\big((k+l)^2+\Lambda^2\big)}\Big).\qquad
\end{eqnarray}

\noindent
The integral $I_2$ can be calculated by the standard methods, see, e.g. \cite{Soloshenko:2002np}. The result is given by the expression

\begin{equation}\label{I2}
I_2 = \frac{1}{8\pi^2}\Big(\ln\frac{\Lambda}{k} + \sqrt{1+ \frac{4\Lambda^2}{k^2}} \mbox{arctanh} \sqrt{\frac{k^2}{k^2+4\Lambda^2}}\Big).
\end{equation}

\noindent
The integral $I_1$ can be calculated by the method similar to the one considered in \cite{Soloshenko:2002np,Soloshenko:2003sx}. Namely, we use the four-dimensional spherical coordinates

\begin{eqnarray}
&& l_1 = l \sin\theta_3\,\sin\theta_2\,\sin\theta_1;\qquad l_2 = l \sin\theta_3\,\sin\theta_2\,\cos\theta_1;\nonumber\\
&& l_3 = l \sin\theta_3\,\cos\theta_2;\qquad\qquad \ \, l_4 = l \cos\theta_3,
\end{eqnarray}

\noindent
in which the integration measure is given by

\begin{equation}
\int d^4l = \int\limits_0^\infty dl\, l^3 \int\limits_0^\pi d\theta_3\,\sin^2\theta_3 \int\limits_0^\pi d\theta_2\,\sin\theta_2 \int\limits_0^{2\pi} d\theta_1.
\end{equation}

\noindent
If the fourth axis is directed collinear to the vector $k_\mu$, then $(k+l)^2 = k^2 +2kl \cos\theta_3 + l^2$, and the integrand in the expression (\ref{I1_Definition}) depends only on $\theta_3$. In this case, after the substitution $x\equiv \cos\theta_3$, the integration measure can be written in the form

\begin{equation}
\int d^4l \to 4\pi \int\limits_0^\infty dl\, l^3 \int\limits_{-1}^1 dx\sqrt{1-x^2}.
\end{equation}

\begin{figure}[h]
\begin{picture}(0,4)
\put(5.6,0){\includegraphics[scale=0.29]{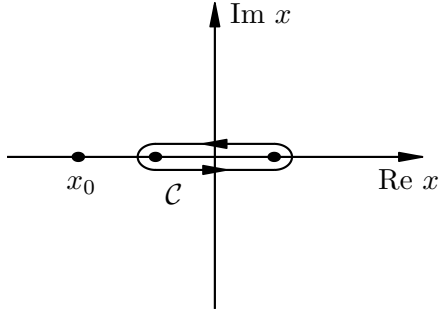}}
\put(6.4,1.6){$x_0$}
\put(7.7,1.4){${\cal C}$}
\put(10.5,1.6){Re $x$}
\put(8.55,3.8){Im $x$}
\end{picture}
\caption{The contour ${\cal C}$ in the $x$ complex plane which is used for integrating over the angle $\theta_3$.}\label{Figure_Contour}
\end{figure}

\noindent
Consequently, the integral $I_1$ can be presented as

\begin{equation}
I_1 = \frac{1}{16\pi^3}\int\limits_{0}^\infty dl^2\, \frac{\Lambda^2}{l^2+\Lambda^2} \oint\limits_{\cal C} dx\,\frac{\sqrt{1-x^2}}{k^2+2klx+l^2},
\end{equation}

\noindent
where ${\cal C}$ is the contour in the complex $x$-plane shown in Fig. \ref{Figure_Contour}. The contour integral can be found by calculating the residues at infinity and at $x_0 = -(k^2+l^2)/2kl$, see Ref. \cite{Soloshenko:2002np,Soloshenko:2003sx} for details. The result is written as

\begin{equation}
\oint\limits_{\cal C} dx\, \frac{\sqrt{1-x^2}}{k^2+2klx+l^2} = \left\{\begin{array}{l} {\displaystyle \frac{\pi}{k^2}\qquad\mbox{for}\quad k\ge l}\\
\vphantom{1}\\
\displaystyle{\frac{\pi}{l^2}\qquad\mbox{for}\quad l\ge k.}\end{array}\right.
\end{equation}

\noindent
Using this equation it is possible to calculate the angular part of the integral $I_1$, so that

\begin{eqnarray}\label{I1}
&& I_1 = \frac{1}{16\pi^2} \int\limits_{0}^{k^2} dl^2 \frac{\Lambda^2}{k^2(l^2+\Lambda^2)} + \frac{1}{16\pi^2} \int\limits_{k^2}^\infty dl^2 \frac{\Lambda^2}{l^2(l^2+\Lambda^2)}\nonumber\\
&&\qquad\qquad\qquad\qquad\qquad = \frac{\Lambda^2}{16\pi^2 k^2} \ln\Big(1+\frac{k^2}{\Lambda^2}\Big) + \frac{1}{16\pi^2}\ln\Big(1+\frac{\Lambda^2}{k^2}\Big).\qquad
\end{eqnarray}

\noindent
From Eqs. (\ref{I1}) and (\ref{I2}) we obtain

\begin{equation}
2I_1-I_2 = \frac{1}{8\pi^2} \ln\frac{\Lambda}{k}+\frac{1}{8\pi^2}\Big(1+\frac{\Lambda^2}{k^2}\Big) \ln\Big(1+\frac{k^2}{\Lambda^2}\Big) - \frac{1}{8\pi^2}\sqrt{1+ \frac{4\Lambda^2}{k^2}} \mbox{arctanh} \sqrt{\frac{k^2}{k^2+4\Lambda^2}}.
\end{equation}

\noindent
Thus, the expression (\ref{Integral_We_Calculate}) can be presented in the form

\begin{eqnarray}\label{Last_Integral}
&& I = \frac{1}{8\pi^2} \frac{d}{d\ln\Lambda} \int \frac{d^4k}{(2\pi)^4} \frac{1}{k^4 F_k^3}\Bigg\{\ln\frac{\Lambda}{k} - F_k \Big(\ln\frac{\Lambda}{\mu} + g_1\Big)\nonumber\\
&&\qquad\qquad\qquad + \Big(1+\frac{\Lambda^2}{k^2}\Big) \ln\Big(1+\frac{k^2}{\Lambda^2}\Big) - \sqrt{1+ \frac{4\Lambda^2}{k^2}} \mbox{arctanh} \sqrt{\frac{k^2}{k^2+4\Lambda^2}} \Bigg\}.\qquad
\end{eqnarray}

\noindent
Let us note that contribution of the last two terms in the brackets vanishes,

\begin{equation}\label{Vanishing_Integral}
\frac{d}{d\ln\Lambda} \int \frac{d^4k}{(2\pi)^4} \frac{1}{k^4 F_k^3}\Bigg\{\Big(1+\frac{\Lambda^2}{k^2}\Big) \ln\Big(1+\frac{k^2}{\Lambda^2}\Big) - \sqrt{1+ \frac{4\Lambda^2}{k^2}} \mbox{arctanh} \sqrt{\frac{k^2}{k^2+4\Lambda^2}} \Bigg\} = 0.
\end{equation}

\noindent
Really, the expression in the large brackets rapidly tends to 0 in the limit $k\to 0$, while the function $F_k$ rapidly increases at infinity. This implies that the integral in Eq. (\ref{Vanishing_Integral}) is convergent. Consequently, the dependence on $\Lambda$ can be eliminated by the substitution $k_\mu = \Lambda K_\mu$. Therefore, the considered integral is independent of $\Lambda$, and its derivative with respect to $\ln\Lambda$ vanishes.

Then, we proceed to calculating the remaining part of the expression (\ref{Last_Integral}). It should be noted that the integral of the first two terms is not well-defined, because it diverges at $k=0$. However, the derivative with respect to $\ln\Lambda$ eliminates this problem, if we perform the integration over $d^4k$ after the differentiation. After differentiating with respect to $\ln\Lambda$ we obtain

\begin{eqnarray}\label{Final_Integral}
&& I = \frac{1}{8\pi^2} \frac{d}{d\ln\Lambda} \int \frac{d^4k}{(2\pi)^4} \frac{1}{k^4 F_k^3}\Big\{\ln\frac{\Lambda}{k} - F_k \Big(\ln\frac{\Lambda}{\mu} + g_1\Big)\Big\}\nonumber\\
&& = \frac{1}{8\pi^2} \int \frac{d^4k}{(2\pi)^4} \frac{1}{k^4} \Big\{\frac{(1-F_k)}{F_k^3} - \Big(\ln\frac{\Lambda}{\mu}+g_1\Big) \frac{d}{d\ln\Lambda}\Big(\frac{1}{F_k^2}\Big) + \ln\frac{\Lambda}{k}\, \frac{d}{d\ln\Lambda}\Big(\frac{1}{F_k^3}\Big) \Big\}.\qquad
\end{eqnarray}

\noindent
The integral corresponding to the first term in the brackets can be calculated straightforwardly in the four-dimensional spherical coordinates taking into account that the volume of the unit sphere $S^3$ is $2\pi^2$,

\begin{equation}\label{Term1}
\int \frac{d^4k}{(2\pi)^4} \frac{(1-F_k)}{k^4 F_k^3} = -\frac{\Lambda^4}{16\pi^2} \int\limits_{0}^\infty \frac{dk^2}{(k^2+\Lambda^2)^3} = - \frac{1}{32\pi^2}.
\end{equation}

\noindent
To find a contribution of the second term in Eq. (\ref{Final_Integral}), we note that the function $F_k$ depends on $k/\Lambda$, so that the derivative with respect to $\ln\Lambda$ can be converted into the derivative with respect to $\ln k$,

\begin{equation}\label{Term2}
\int \frac{d^4k}{(2\pi)^4}\,\frac{1}{k^4}\,\frac{d}{d\ln\Lambda}\Big(\frac{1}{F_k^2}\Big) = -\frac{1}{8\pi^2} \int\limits_0^\infty dk\,\frac{d}{dk} \Big(\frac{1}{F_k^2}\Big) = \frac{1}{8\pi^2 F_k^2(k=0)} = \frac{1}{8\pi^2}.
\end{equation}

\noindent
The contribution of the last term in Eq. (\ref{Final_Integral}) in the four-dimensional spherical coordinates takes the form

\begin{equation}\label{Term3}
\int \frac{d^4k}{(2\pi)^4}\,\frac{1}{k^4}\ln\frac{\Lambda}{k}\,\frac{d}{d\ln\Lambda}\Big(\frac{1}{F_k^3}\Big) = -\frac{1}{8\pi^2} \int\limits_0^\infty dk\,\ln\frac{\Lambda}{k}\,\frac{d}{dk} \Big(\frac{1}{F_k^3}\Big).
\end{equation}

\noindent
It is easy to see that this integral is convergent both at infinity and at $k=0$. (The derivative of $F_k$ with respect to $\ln\Lambda$ is proportional to $k^2$ in the limit $k\to 0$.) Therefore, it is possible to replace the lower integration limit by $\varepsilon\to 0$. After this, integrating by parts we obtain

\begin{eqnarray}
&& -\frac{1}{8\pi^2 F_k^3} \ln\frac{\Lambda}{k}\,\Bigg|_{\varepsilon}^\infty -\frac{1}{8\pi^2} \int\limits_\varepsilon^\infty \frac{dk}{k F_k^3} = \frac{1}{8\pi^2} \ln\frac{\Lambda}{\varepsilon} - \frac{\Lambda^6}{16\pi^2} \int\limits_\varepsilon^\infty \frac{dk^2}{k^2 (k^2+\Lambda^2)^3}\nonumber\\
&& =\frac{1}{8\pi^2} \ln\frac{\Lambda}{\varepsilon} - \frac{1}{16\pi^2}\Bigg(\frac{\Lambda^4}{2(k^2+\Lambda^2)^2} + \frac{\Lambda^2}{k^2+\Lambda^2} - \ln\Big(1+\frac{\Lambda^2}{k^2}\Big) \Bigg)\Bigg|_{\varepsilon}^\infty = \frac{3}{32\pi^2}.
\end{eqnarray}

\noindent
Using Eqs. (\ref{Term1}), (\ref{Term2}), and (\ref{Term3}) we find the result for the integral (\ref{Integral_We_Calculate}),

\begin{equation}
I = \frac{1}{64\pi^4}\Big[\,\frac{1}{2} - \Big(\ln\frac{\Lambda}{\mu}+g_1\Big)\Big].
\end{equation}

%%%%%%%%%%%%%%%%%%%%%%%%%%%%%%%%%%%%%%%%%%%%%%%%%%%%%%%%%%%%%%%%%%%%%%%%%


\begin{thebibliography}{100}

%\cite{Novikov:1983uc}
\bibitem{Novikov:1983uc}
  V.~A.~Novikov, M.~A.~Shifman, A.~I.~Vainshtein and V.~I.~Zakharov,
  %``Exact Gell-Mann-Low Function of Supersymmetric Yang-Mills Theories from Instanton Calculus,''
  Nucl.\ Phys.\ B {\bf 229} (1983) 381.
  %%CITATION = NUPHA,B229,381;%%

%\cite{Jones:1983ip}
\bibitem{Jones:1983ip}
  D.~R.~T.~Jones,
  %``More on the Axial Anomaly in Supersymmetric {Yang-Mills} Theory,''
  Phys.\ Lett.\ B {\bf 123} (1983) 45.
  %%CITATION = PHLTA,B123,45;%%

%\cite{Novikov:1985rd}
\bibitem{Novikov:1985rd}
  V.~A.~Novikov, M.~A.~Shifman, A.~I.~Vainshtein and V.~I.~Zakharov,
  %``Beta Function in Supersymmetric Gauge Theories: Instantons Versus Traditional Approach,''
  Phys.\ Lett.\ B {\bf 166} (1986) 329; Sov.\ J.\ Nucl.\ Phys.\  {\bf 43}
(1986) 294; [Yad.\ Fiz.\  {\bf 43} (1986) 459].
  %%CITATION = PHLTA,B166,329;%%

%\cite{Shifman:1986zi}
\bibitem{Shifman:1986zi}
  M.~A.~Shifman and A.~I.~Vainshtein,
  %``Solution of the Anomaly Puzzle in SUSY Gauge Theories and the Wilson Operator Expansion,''
  Nucl.\ Phys.\ B {\bf 277}  (1986) 456; Sov.\ Phys.\ JETP {\bf 64} (1986) 428;
[Zh.\ Eksp.\ Teor.\ Fiz.\  {\bf 91}  (1986) 723].
  %%CITATION = NUPHA,B277,456;%%

%\cite{Shifman:1999mv}
\bibitem{Shifman:1999mv}
  M.~A.~Shifman and A.~I.~Vainshtein,
  %``Instantons versus supersymmetry: Fifteen years later,''
  In *Shifman, M.A.: ITEP lectures on particle physics and field theory, vol. 2* 485-647
  [hep-th/9902018].
  %%CITATION = HEP-TH/9902018;%%

%\cite{Buchbinder:2014wra}
\bibitem{Buchbinder:2014wra}
  I.~L.~Buchbinder and K.~V.~Stepanyantz,
  %``The higher derivative regularization and quantum corrections in N=2 supersymmetric theories,''
  Nucl.\ Phys.\ B {\bf 883} (2014) 20.
  %[arXiv:1402.5309 [hep-th]].
  %%CITATION = ARXIV:1402.5309;%%

%\cite{Buchbinder:2015eva}
\bibitem{Buchbinder:2015eva}
  I.~L.~Buchbinder, N.~G.~Pletnev and K.~V.~Stepanyantz,
  %``Manifestly N=2 supersymmetric regularization for N=2 supersymmetric field theories,''
  Phys.\ Lett.\ B {\bf 751} (2015) 434.
  %doi:10.1016/j.physletb.2015.10.071
  %[arXiv:1509.08055 [hep-th]].

%\cite{Grisaru:1982zh}
\bibitem{Grisaru:1982zh}
  M.~T.~Grisaru and W.~Siegel,
  %``Supergraphity. 2. Manifestly Covariant Rules and Higher Loop Finiteness,''
  Nucl.\ Phys.\ B {\bf 201} (1982) 292
   [Erratum-ibid.\ B {\bf 206} (1982) 496].
  %%CITATION = NUPHA,B201,292;%%

%\cite{Howe:1983sr}
\bibitem{Howe:1983sr}
  P.~S.~Howe, K.~S.~Stelle and P.~K.~Townsend,
  %``Miraculous Ultraviolet Cancellations in Supersymmetry Made Manifest,''
  Nucl.\ Phys.\ B {\bf 236} (1984) 125.
  %%CITATION = NUPHA,B236,125;%%

%\cite{Buchbinder:1997ib}
\bibitem{Buchbinder:1997ib}
  I.~L.~Buchbinder, S.~M.~Kuzenko and B.~A.~Ovrut,
  %``On the D = 4, N=2 nonrenormalization theorem,''
  Phys.\ Lett.\ B {\bf 433} (1998) 335.
  %[hep-th/9710142].
  %%CITATION = HEP-TH/9710142;%%

%\cite{Hisano:1997ua}
\bibitem{Hisano:1997ua}
  J.~Hisano and M.~A.~Shifman,
  %``Exact results for soft supersymmetry breaking parameters in supersymmetric gauge theories,''
  Phys.\ Rev.\ D {\bf 56} (1997) 5475.
  %doi:10.1103/PhysRevD.56.5475
  %[hep-ph/9705417].
  %%CITATION = doi:10.1103/PhysRevD.56.5475;%%

%\cite{Jack:1997pa}
\bibitem{Jack:1997pa}
  I.~Jack and D.~R.~T.~Jones,
  %``The Gaugino Beta function,''
  Phys.\ Lett.\ B {\bf 415} (1997) 383.
  %doi:10.1016/S0370-2693(97)01277-X
  %[hep-ph/9709364].
  %%CITATION = doi:10.1016/S0370-2693(97)01277-X;%%

%\cite{Avdeev:1997vx}
\bibitem{Avdeev:1997vx}
  L.~V.~Avdeev, D.~I.~Kazakov and I.~N.~Kondrashuk,
  %``Renormalizations in softly broken SUSY gauge theories,''
  Nucl.\ Phys.\ B {\bf 510} (1998) 289.
  %doi:10.1016/S0550-3213(98)81015-8, 10.1016/S0550-3213(97)00706-2
  %[hep-ph/9709397].
  %%CITATION = doi:10.1016/S0550-3213(98)81015-8, 10.1016/S0550-3213(97)00706-2;%%

%\cite{ArkaniHamed:1997mj}
\bibitem{ArkaniHamed:1997mj}
  N.~Arkani-Hamed and H.~Murayama,
  %``Holomorphy, rescaling anomalies and exact beta functions in supersymmetric gauge theories,''
  JHEP {\bf 0006} (2000) 030.
  %[hep-th/9707133].
  %%CITATION = HEP-TH/9707133;%%

%\cite{Kraus:2002nu}
\bibitem{Kraus:2002nu}
  E.~Kraus, C.~Rupp and K.~Sibold,
  %``Supersymmetric Yang-Mills theories with local coupling: The Supersymmetric gauge,''
  Nucl.\ Phys.\ B {\bf 661} (2003) 83.
  %[hep-th/0212064].
  %%CITATION = HEP-TH/0212064;%%

%\cite{Jack:1996vg}
\bibitem{Jack:1996vg}
  I.~Jack, D.~R.~T.~Jones and C.~G.~North,
  %``N=1 supersymmetry and the three loop gauge Beta function,''
  Phys.\ Lett.\ B {\bf 386} (1996) 138.

%\cite{Jack:1996cn}
\bibitem{Jack:1996cn}
  I.~Jack, D.~R.~T.~Jones and C.~G.~North,
  %``Scheme dependence and the NSVZ Beta function,''
  Nucl.\ Phys.\ B {\bf 486} (1997) 479.
  %%CITATION = HEP-PH/9609325;%%

%\cite{Jack:1998uj}
\bibitem{Jack:1998uj}
  I.~Jack, D.~R.~T.~Jones and A.~Pickering,
  %``The Connection between DRED and NSVZ,''
  Phys.\ Lett.\ B {\bf 435} (1998) 61.
  %%CITATION = HEP-PH/9805482;%%

%\cite{Kataev:2013csa}
\bibitem{Kataev:2013csa}
  A.~L.~Kataev and K.~V.~Stepanyantz,
  %``Scheme independent consequence of the NSVZ relation for $N=1$ SQED with $N_f$ flavors,''
  Phys.\ Lett.\ B {\bf 730} (2014) 184.
  %[arXiv:1311.0589 [hep-th]].
  %%CITATION = ARXIV:1311.0589;%%

%\cite{Kutasov:2004xu}
\bibitem{Kutasov:2004xu}
  D.~Kutasov and A.~Schwimmer,
  %``Lagrange multipliers and couplings in supersymmetric field theory,''
  Nucl.\ Phys.\ B {\bf 702} (2004) 369.
  %[hep-th/0409029].
  %%CITATION = HEP-TH/0409029;%%

%\cite{Kataev:2014gxa}
\bibitem{Kataev:2014gxa}
  A.~L.~Kataev and K.~V.~Stepanyantz,
  %``The NSVZ $\beta$-function in supersymmetric theories with different regularizations and renormalization prescriptions,''
  Theor.\ Math.\ Phys.\  {\bf 181} (2014) 1531.
  %[arXiv:1405.7598 [hep-th]].
  %%CITATION = ARXIV:1405.7598;%%

%\cite{Aleshin:2016rrr}
\bibitem{Aleshin:2016rrr}
  S.~S.~Aleshin, I.~O.~Goriachuk, A.~L.~Kataev and K.~V.~Stepanyantz,
  %``The NSVZ scheme for ${\cal N}=1$ SQED with $N_f$ flavors, regularized by the dimensional reduction, in the three-loop approximation,''
  Phys.\ Lett.\ B {\bf 764} (2017) 222.
  %doi:10.1016/j.physletb.2016.11.041
  %[arXiv:1610.08034 [hep-th]].
  %%CITATION = doi:10.1016/j.physletb.2016.11.041;%%

%\cite{Kataev:2013eta}
\bibitem{Kataev:2013eta}
  A.~L.~Kataev and K.~V.~Stepanyantz,
  %``NSVZ scheme with the higher derivative regularization for $\mathcal{N} =$ 1 SQED,''
  Nucl.\ Phys.\ B {\bf 875} (2013) 459.
  %%CITATION = ARXIV:1305.7094;%%

%\cite{Slavnov:1971aw}
\bibitem{Slavnov:1971aw}
  A.~A.~Slavnov,
  %``Invariant regularization of nonlinear chiral theories,''
  Nucl.\ Phys.\ B {\bf 31} (1971) 301.
  %%CITATION = NUPHA,B31,301;%%

%\cite{Slavnov:1972sq}
\bibitem{Slavnov:1972sq}
  A.~A.~Slavnov,
  %``Invariant regularization of gauge theories,''
  Theor.Math.Phys. {\bf 13} (1972) 1064
   [Teor.\ Mat.\ Fiz.\  {\bf 13} (1972) 174].
  %%CITATION = TMFZA,13,174;%%

%\cite{Slavnov:1977zf}
\bibitem{Slavnov:1977zf}
  A.~A.~Slavnov,
  %``The Pauli-Villars Regularization for Nonabelian Gauge Theories,''
  Theor.\ Math.\ Phys.\ {\bf 33} (1977) 977
   [Teor.\ Mat.\ Fiz.\  {\bf 33} (1977) 210].
  %%CITATION = TMFZA,33,210;%%

%\cite{Krivoshchekov:1978xg}
\bibitem{Krivoshchekov:1978xg}
  V.~K.~Krivoshchekov,
  %``Invariant Regularizations for Supersymmetric Gauge Theories,''
  Theor.\ Math.\ Phys.\ {\bf 36} (1978) 745
 [Teor.\ Mat.\ Fiz.\  {\bf 36} (1978) 291].
 %%CITATION = TMFZA,36,291;%%

%\cite{West:1985jx}
\bibitem{West:1985jx}
  P.~C.~West,
  %``Higher Derivative Regulation Of Supersymmetric Theories,''
  Nucl.\ Phys.\ B {\bf 268} (1986) 113.
  %%CITATION = NUPHA,B268,113;%%

%\cite{Siegel:1979wq}
\bibitem{Siegel:1979wq}
  W.~Siegel,
  %``Supersymmetric Dimensional Regularization via Dimensional Reduction,''
  Phys.\ Lett.\ B {\bf 84} (1979) 193.
  %%CITATION = PHLTA,B84,193;%%

%\cite{Aleshin:2015qqc}
\bibitem{Aleshin:2015qqc}
  S.~S.~Aleshin, A.~L.~Kataev and K.~V.~Stepanyantz,
  %``Structure of three-loop contributions to the $\beta$-function of $\mathcal N = 1$ supersymmetric QED with N$_{f}$ flavors regularized by the dimensional reduction,''
  JETP Lett.\  {\bf 103} (2016) no.2,  77
  [Pisma Zh.\ Eksp.\ Teor.\ Fiz.\  {\bf 103} (2016) 83].
  %doi:10.1134/S0021364016020028
  %[arXiv:1511.05675 [hep-th]].
  %%CITATION = doi:10.1134/S0021364016020028;%%

%\cite{Soloshenko:2003nc}
\bibitem{Soloshenko:2003nc}
  A.~A.~Soloshenko and K.~V.~Stepanyantz,
  %``Three loop beta function for N=1 supersymmetric electrodynamics, regularized by higher derivatives,''
  Theor.\ Math.\ Phys.\  {\bf 140} (2004) 1264
   [Teor.\ Mat.\ Fiz.\  {\bf 140} (2004) 430].
  %%CITATION = HEP-TH/0304083;%%

%\cite{Smilga:2004zr}
\bibitem{Smilga:2004zr}
  A.~V.~Smilga and A.~Vainshtein,
  %``Background field calculations and nonrenormalization theorems in 4-D supersymmetric gauge theories and their low-dimensional descendants,''
  Nucl.\ Phys.\ B {\bf 704} (2005) 445.
  %%CITATION = HEP-TH/0405142;%%

\bibitem{Shevtsova2009}
E.~S.~Shevtsova and K.~V.~Stepanyantz,
Moscow\ Univ.\ Phys.\ Bull.\ {\bf 64} (2009) 485.

%\cite{Pimenov:2009hv}
\bibitem{Pimenov:2009hv}
  A.~B.~Pimenov, E.~S.~Shevtsova and K.~V.~Stepanyantz,
  %``Calculation of two-loop beta-function for general N=1 supersymmetric Yang--Mills theory with the higher covariant derivative regularization,''
  Phys.\ Lett.\ B {\bf 686} (2010) 293.
  %%CITATION = ARXIV:0912.5191;%%

%\cite{Stepanyantz:2011bz}
\bibitem{Stepanyantz:2011bz}
  K.~V.~Stepanyantz,
  ``Factorization of integrals defining the two-loop $\beta$-function for the general renormalizable N=1 SYM theory, regularized by the higher covariant derivatives, into integrals of double total derivatives,''
  arXiv:1108.1491 [hep-th].
  %%CITATION = ARXIV:1108.1491;%%

%\cite{Stepanyantz:2012zz}
\bibitem{Stepanyantz:2012zz}
  K.~V.~Stepanyantz,
  %``Derivation of the exact NSVZ beta-function in N=1 SQED regularized by higher derivatives by summation of Feynman diagrams,''
  J.\ Phys.\ Conf.\ Ser.\  {\bf 343} (2012) 012115.
  %%CITATION = 00462,343,012115;%%

%\cite{Stepanyantz:2012us}
\bibitem{Stepanyantz:2012us}
  K.~V.~Stepanyantz,
  %``Multiloop calculations in supersymmetric theories with the higher covariant derivative regularization,''
  J.\ Phys.\ Conf.\ Ser.\  {\bf 368} (2012) 012052.
  %%CITATION = ARXIV:1203.5525;%%

%\cite{Kazantsev:2014yna}
\bibitem{Kazantsev:2014yna}
  A.~E.~Kazantsev and K.~V.~Stepanyantz,
  %``Relation between two-point Green functions of ${\cal N}=1$ SQED with $N_f$ flavors, regularized by higher derivatives, in the three-loop approximation,''
  J.\ Exp.\ Theor.\ Phys. {\bf 120} (2015) 618 [Zh.\ Eksp.\ Teor.\ Fiz. {\bf 147} (2015) 714].
  %[arXiv:1410.1133 [hep-th]].
  %%CITATION = ARXIV:1410.1133;%%

%\cite{Stepanyantz:2011jy}
\bibitem{Stepanyantz:2011jy}
  K.~V.~Stepanyantz,
  %``Derivation of the exact NSVZ $\beta$-function in N=1 SQED, regularized by higher derivatives, by direct summation of Feynman diagrams,''
  Nucl.\ Phys.\ B {\bf 852} (2011) 71.
  %%CITATION = ARXIV:1102.3772;%%

%\cite{Stepanyantz:2014ima}
\bibitem{Stepanyantz:2014ima}
  K.~V.~Stepanyantz,
  %``The NSVZ $\beta$-function and the Schwinger-Dyson equations for $\mathcal{N}=1$ SQED with $N_{f}$ flavors, regularized by higher derivatives,''
  JHEP {\bf 1408} (2014) 096.
  %[arXiv:1404.6717 [hep-th]].
  %%CITATION = ARXIV:1404.6717;%%

%\cite{Adler:1974gd}
\bibitem{Adler:1974gd}
  S.~L.~Adler,
  %``Some Simple Vacuum Polarization Phenomenology: e+ e- ---> Hadrons: The mu - Mesic Atom x-Ray Discrepancy and (g-2) of the Muon,''
  Phys.\ Rev.\ D {\bf 10} (1974) 3714.
  %%CITATION = PHRVA,D10,3714;%%

%\cite{Shifman:2014cya}
\bibitem{Shifman:2014cya}
  M.~Shifman and K.~Stepanyantz,
  %*``Exact Adler Function in Supersymmetric QCD,''
  Phys.\ Rev.\ Lett.\  {\bf 114} (2015) 051601.
  %[arXiv:1412.3382 [hep-th]].
  %%CITATION = ARXIV:1412.3382;%%

%\cite{Shifman:2015doa}
\bibitem{Shifman:2015doa}
  M.~Shifman and K.~V.~Stepanyantz,
  %``Derivation of the exact expression for the D function in N=1 SQCD,''
  Phys.\ Rev.\ D {\bf 91} (2015) 105008.
  %[arXiv:1502.06655 [hep-th]].

%\cite{Nartsev:2016nym}
\bibitem{Nartsev:2016nym}
  I.~V.~Nartsev and K.~V.~Stepanyantz,
  ``Exact renormalization of the photino mass in softly broken ${\cal N}=1$ SQED with $N_f$ flavors regularized by higher derivatives,''
  arXiv:1610.01280 [hep-th].
  %%CITATION = ARXIV:1610.01280;%%

%\cite{Bogolyubov:1980nc}
\bibitem{Bogolyubov:1980nc}
  N.~N.~Bogolyubov and D.~V.~Shirkov,
  ``Introduction To The Theory Of Quantized Fields,''
  Nauka, Moscow, 1984
  [Intersci.\ Monogr.\ Phys.\ Astron.\  {\bf 3} (1959) 1].
  %%CITATION = IMTPA,3,1;%%

%\cite{Nartsev:2016mvn}
\bibitem{Nartsev:2016mvn}
  I.~V.~Nartsev and K.~V.~Stepanyantz,
  %``NSVZ-like scheme for the photino mass in softly broken ${\cal N}=1$ SQED regularized by higher derivatives,''
  JETP Lett.\  {\bf 105} (2017) no.2,  69
  [Pisma Zh.\ Eksp.\ Teor.\ Fiz.\  {\bf 105} (2017) 57].
  %[arXiv:1611.09091 [hep-th]].
  %%CITATION = ARXIV:1611.09091;%%

%\cite{Stepanyantz:2016gtk}
\bibitem{Stepanyantz:2016gtk}
  K.~V.~Stepanyantz,
  %``Non-renormalization of the $V\bar cc$-vertices in ${\cal N}=1$ supersymmetric theories,''
  Nucl.\ Phys.\ B {\bf 909} (2016) 316.
  %doi:10.1016/j.nuclphysb.2016.05.011
  %[arXiv:1603.04801 [hep-th]].
  %%CITATION = doi:10.1016/j.nuclphysb.2016.05.011;%%

%\cite{West:1990tg}
\bibitem{West:1990tg}
  P.~C.~West,
  ``Introduction to supersymmetry and supergravity,''
  Singapore, Singapore: World Scientific (1990) 425 p.

%\cite{Buchbinder:1998qv}
\bibitem{Buchbinder:1998qv}
  I.~L.~Buchbinder and S.~M.~Kuzenko,
  ``Ideas and methods of supersymmetry and supergravity: Or a walk through superspace,''
  Bristol, UK: IOP (1998) 656 p.

%\cite{Aleshin:2016yvj}
\bibitem{Aleshin:2016yvj}
  S.~S.~Aleshin, A.~E.~Kazantsev, M.~B.~Skoptsov and K.~V.~Stepanyantz,
  %``One-loop divergences in non-Abelian supersymmetric theories regularized by BRST-invariant version of the higher derivative regularization,''
  JHEP {\bf 1605} (2016) 014.
  %doi:10.1007/JHEP05(2016)014
  %[arXiv:1603.04347 [hep-th]].
  %%CITATION = doi:10.1007/JHEP05(2016)014;%%

%\cite{Grisaru:1979wc}
\bibitem{Grisaru:1979wc}
  M.~T.~Grisaru, W.~Siegel and M.~Rocek,
  %``Improved Methods for Supergraphs,''
  Nucl.\ Phys.\ B {\bf 159} (1979) 429.
  %%CITATION = NUPHA,B159,429;%%

%\cite{Soloshenko:2002np}
\bibitem{Soloshenko:2002np}
  A.~Soloshenko and K.~Stepanyantz,
  ``Two loop renormalization of N=1 supersymmetric electrodynamics, regularized by higher derivatives,''
  hep-th/0203118.
  %%CITATION = HEP-TH/0203118;%%

%\cite{Soloshenko:2003sx}
\bibitem{Soloshenko:2003sx}
  A.~A.~Soloshenko and K.~V.~Stepanyantz,
  %``Two-loop anomalous dimension of N = 1 supersymmetric quantum electrodynamics regularized using higher covariant derivatives,''
  Theor.\ Math.\ Phys.\  {\bf 134} (2003) 377
   [Teor.\ Mat.\ Fiz.\  {\bf 134} (2003) 430].
  %doi:10.1023/A:1022653506397
  %%CITATION = doi:10.1023/A:1022653506397;%%

\end{thebibliography}
\end{document}